\documentclass[%
reprint,
nofootinbib,
amsmath,amssymb,
aps,
pre,
]{revtex4-1}
\usepackage{lipsum}
\usepackage{xcolor}
\usepackage{amsmath,amssymb,bbm}
\usepackage{soul}
\usepackage{graphicx}
\usepackage{dcolumn}
\usepackage{bm}
\usepackage[colorlinks=true, citecolor=blue]{hyperref}
\usepackage[mathlines]{lineno}
\usepackage[xcolor,framemethod=TikZ]{mdframed}
\mdfdefinestyle{myStyle}{roundcorner=5pt,backgroundcolor=blue!75!cyan!05,linecolor=blue!40!black,linewidth=1.5pt}
\usepackage{comment}

\usepackage{algorithmicx}
\usepackage[ruled,lined]{algorithm2e}
\usepackage{algpseudocode}

\usepackage{stackengine}
\usepackage{float}

\usepackage{scalerel}
\usepackage{tikz}
\usetikzlibrary{svg.path}
\usepackage{stmaryrd}
\usepackage{enumitem}
\usepackage{dblfloatfix}
\setlist[itemize]{align=parleft,left=0pt..1em}

\usepackage[font=footnotesize,caption=false]{subfig}

\newcommand{\dd}{\mathrm{d}}

\newcommand{\PP}{\mathbb{P}}

\graphicspath{{./Figures_new/}}

\begin{document}
	
	\allowdisplaybreaks
	
	\title{Countering adversarial perturbations in graphs using error correcting codes}
	
\author{Saif Eddin Jabari}
\email{sej7@nyu.edu}
\affiliation{New York University Abu Dhabi, Saadiyat Island, P.O. Box 129188, Abu Dhabi, U.A.E.}
\affiliation{New York University Tandon School of Engineering, Brooklyn NY.}

	
\begin{abstract}
We consider the problem of a graph subjected to adversarial perturbations, such as those arising from cyber-attacks, where edges are covertly added or removed. The adversarial perturbations occur during the transmission of the graph between a sender and a receiver. To counteract potential perturbations, this study explores a repetition coding scheme with sender-assigned noise and majority voting on the receiver's end to rectify the graph's structure. The approach operates without prior knowledge of the attack's characteristics. We analytically derive a bound on the number of repetitions needed to satisfy probabilistic constraints on the quality of the reconstructed graph. The method can accurately and effectively decode Erd\H{o}s-R\'{e}nyi graphs that were subjected to non-random edge removal, namely, those connected to vertices with the highest eigenvector centrality, in addition to random addition and removal of edges by the attacker. The method is also effective against attacks on large scale-free graphs generated using the Barab\'{a}si-Albert model but require a larger number of repetitions than needed to correct Erd\H{o}s-R\'{e}nyi graphs.

\end{abstract}
	
\keywords{Error correcting codes}
\maketitle
	
\onecolumngrid
\begin{figure*}[!h]
\centering
\includegraphics[width=0.49\textwidth]{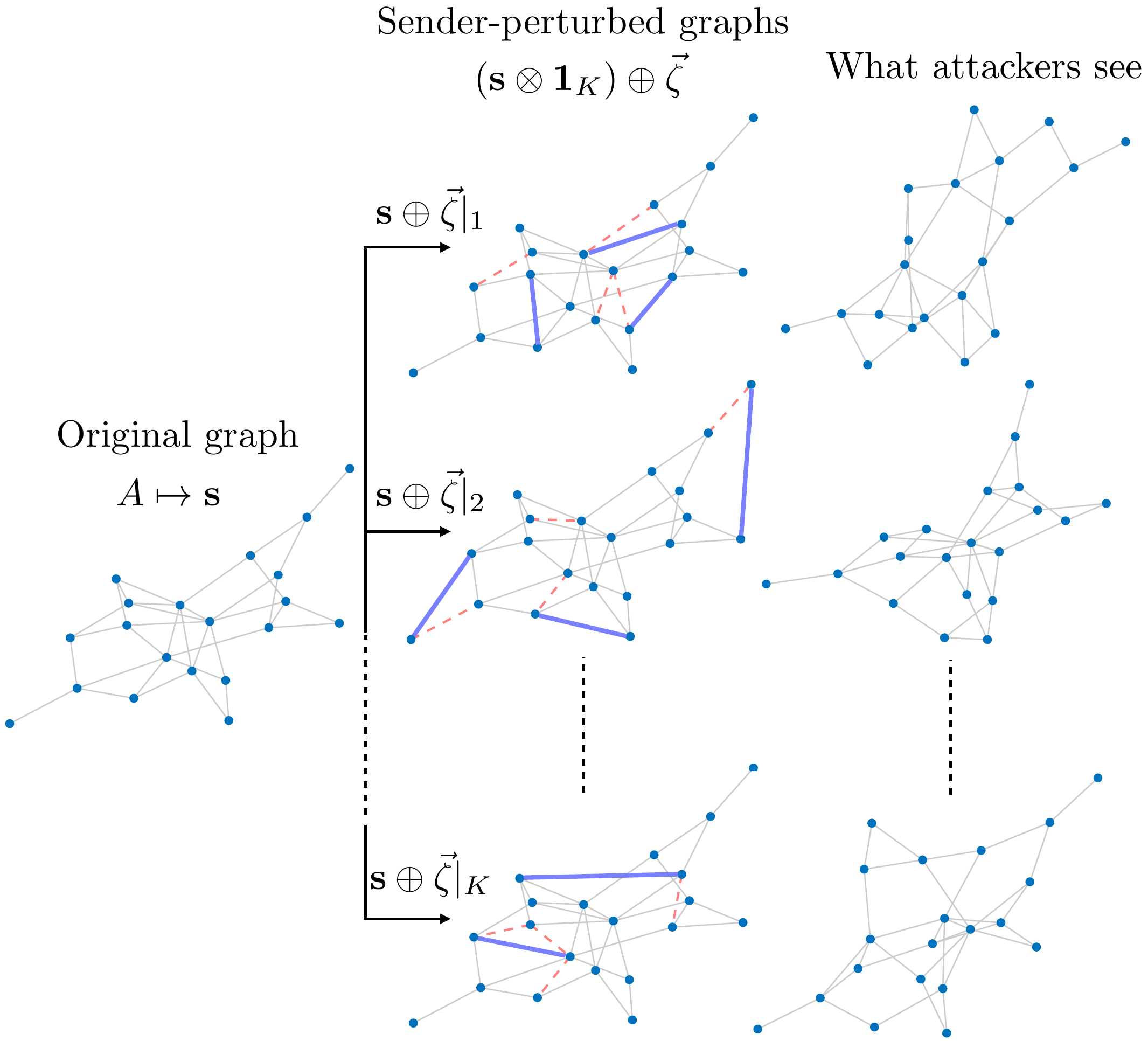} \includegraphics[width=0.49\textwidth]{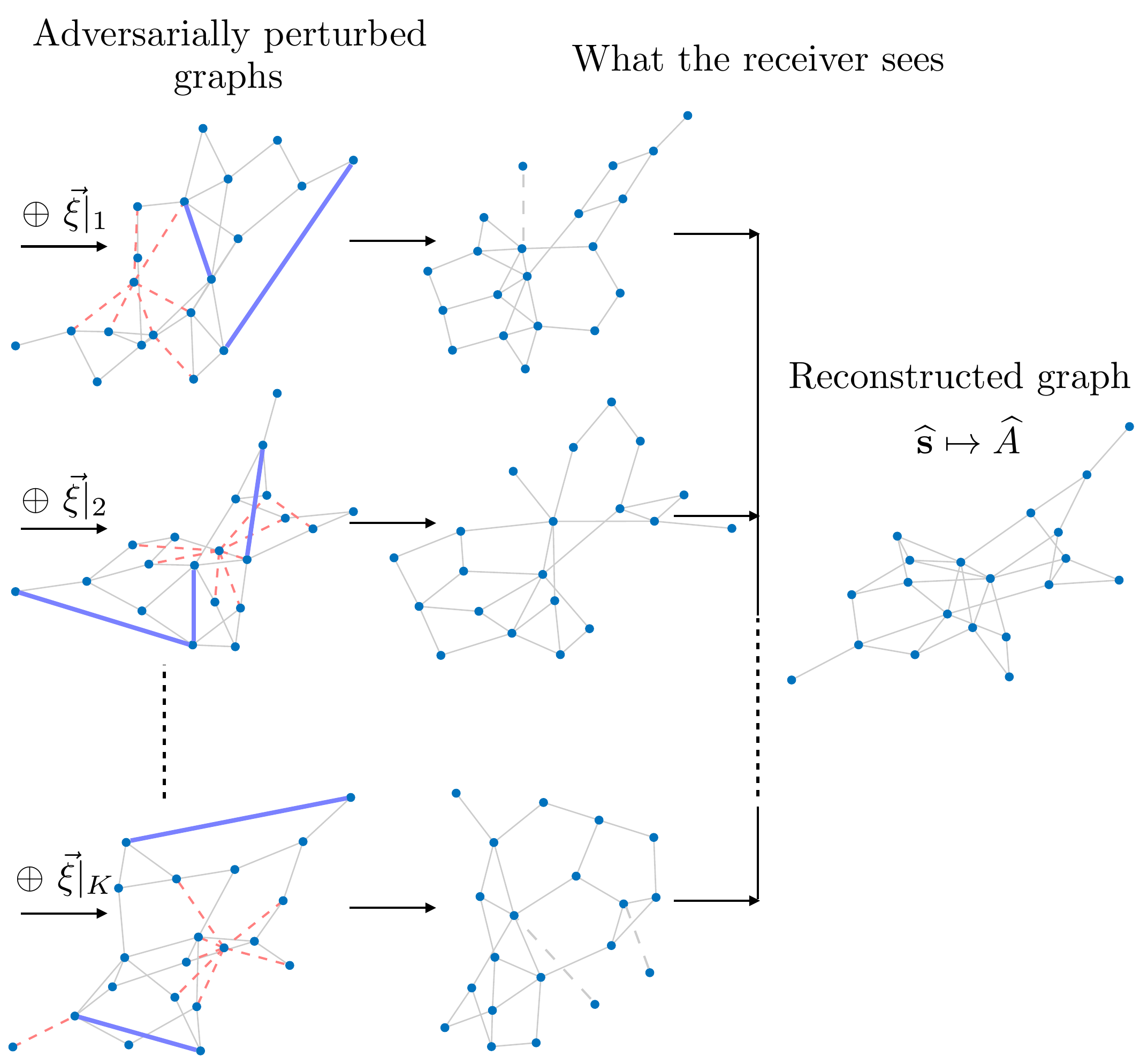}
\caption{An illustration of the proposed graph encoding and decoding schemes. The adjacency matrix $A$ is first vectorized $A \mapsto \mathbf{s}$. $K$ copies of $\mathbf{s}$ are transmitted, each randomly perturbed by the sender $\mathbf{s} \mapsto (\mathbf{s} \otimes \mathbf{1}_K) \oplus \vec{\zeta}$. The $K$ graphs undergo adversarial perturbation by attackers. The nature of the attack is not known to the defenders. In the illustration, the attacker removes all edges attached to the vertex with the highest eigenvector centrality and randomly removes and adds other edges to the graph. Finally, the receiver applies majority voting on the perturbed graphs to reconstruct an estimate of the adjacency matrix $\widehat{A}$. Solid blue edges in the figure are randomly added edges, and dashed red edges are edges removed from the graph.}
\label{fig:general}
\end{figure*}
\twocolumngrid

\section{Introduction}
Machine learning techniques that use deep neural networks, notably graph neural networks (GNNs), have received great attention in various areas of physics \citep{xie2018crystal,huang2021neural,desai2021variational,soltani2022exploring,ehrke2023topological,yu2024spin}.  The interest stems from graphs encoding complex relationships between the constituents of the inputs that traditional vector/matrix representations (resulting in simple Euclidean relationships between the constituents) cannot. This work addresses the problem of correcting adversarial perturbations to graphs, a topic of great importance in defending against cyber-attacks \citep{vivek2019cyberphysical}.
Specifically, we consider an adversarial environment that includes an attacker and a defender. The defender is a GNN that takes graphs as inputs.  The attacker can manipulate the inputs (graphs) of the GNNs by adding or removing edges \cite{zugner2018adversarial,dai2018adversarial,bojchevski2019adversarial,wu2019adversarial,ma2020towards}.  The purpose of such attacks is to trigger misclassification in the GNNs. Studies \cite{zugner2018adversarial,dai2018adversarial,bojchevski2019adversarial,wu2019adversarial,ma2020towards} focused on the attack models and techniques to solve them. \citet{wu2019adversarial} and \citet{zhang2020gnnguard} developed adversarial training methods that defend GNNs against poisoned inputs during training. \citet{gosch2024adversarial} derived theoretical guarantees of robustness for adversarial training methods in prior works. This paper aims to complement these works on adversarial training.

Specifically, this paper provides a mathematical analysis of error-correcting codes used to undo the attack (or de-noise the input graphs). We consider a scenario in which the defender involves two parties, a sender and receiver, where in between, the graph is susceptible to an adversarial attack. The method employs an encoding scheme in which the sender randomly perturbs $K$ copies of the graph, and the receiver applies a majority voting scheme on the $K$ copies received (after being subjected to adversarial perturbations); see Fig.~\ref{fig:general} for illustration. 

We analyze the number of repetitions needed to reliably reconstruct the original graph (Sec.~\ref{sec:analysis}) and derive a mathematical bound on the number of repetitions needed to ensure that the difference between the original and reconstructed graphs falls below a defined threshold with high probability. The number of repetitions needed depends explicitly on the (user-defined) threshold and the graph size. The approach assumes no knowledge of the attack's nature or the original graph's structure, only that less than half the edges are perturbed. Large perturbations are easy to identify, so we focus on imperceptible adversarial perturbations.

The rest of this paper is organized as follows: Sec.~\ref{sec:preliminaries} provides some technical preliminaries that include notation and definitions of GNN robustness from the literature. Sec.~\ref{sec:method} formally presents the encoding and decoding schemes proposed, namely the sender-assigned perturbation and the majority voting scheme. The proposed error-correcting code is analyzed in Sec.~\ref{sec:analysis}, culminating in mathematical bounds on the number of repetitions needed to achieve the probabilistic reliability requirements. Finally, Sec.~\ref{sec:experiments} presents simulation experiments illustrating the effectiveness of the method and Sec.~\ref{sec:conclusion} concludes the paper.

\section{Preliminaries}\label{sec:preliminaries}
\subsection{Notation}
\begin{figure}[h]
\centering
\includegraphics[width=0.4\textwidth]{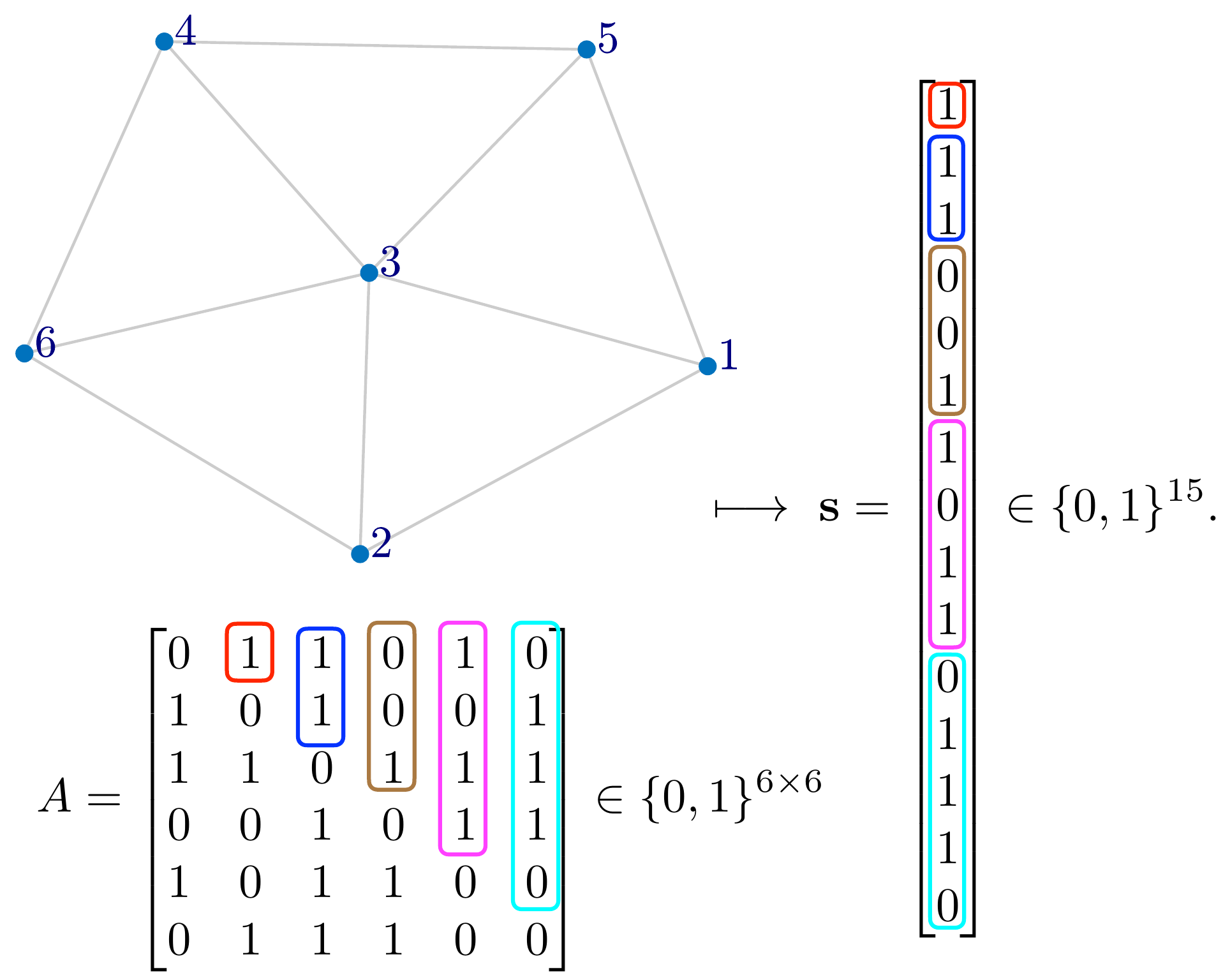}
\caption{An example graph with $|\mathcal{V}| = 6 \Rightarrow N = 15$. Clearly, the relationship between $A$ and $\mathbf{s}$ is one-to-one, that is, given $A$, one calculates a unique vectorization $\mathbf{s}$, and vice versa. The two representations are, thus, equivalent. We use the vector representation to simplify the mathematical exposition.}
\label{fig:EgGraph}
\end{figure}

Let $A \in \{0,1\}^{|\mathcal{V}| \times |\mathcal{V}|}$ denote the adjacency matrix of a graph with vertex set $\mathcal{V}$, which represents an instance of the inputs of a GNN. By vectorizing the off-diagonal elements of $A$, we obtain a flat representation, $\mathbf{s}  \in \{0,1\}^N$, where   $N = \frac{1}{2}(|\mathcal{V}|^2-|\mathcal{V}|)$ is the number of elements in this representation. Fig.~\ref{fig:EgGraph} provides an illustrative example.

In essence, $\mathbf{s}$ vectorizes the upper (equivalently, lower) triangle of $A$. 
A perturbation vector $\vec{\xi} \in\{0,1\}^N$ is introduced to modify the graph structure, with elements representing changes made by an attacker, i.e., $\xi_i=1$ represents the attacker changing the $i$th element of $\mathbf{s}$ and $\xi_i = 0$ leaves the $i$th element unchanged. The perturbation is achieved using the XOR operator, resulting in a perturbed vector. Hence, each element undergoes a specific transformation based on the values of the perturbation vector and the original structure vector.

The attacker can flip any bit in $\mathbf{s}$ (representing an edge or the absence thereof). The perturbation mechanism is only known to the attacker as long as the percentage of edges perturbed is less than one-half. Otherwise, the attack would be easy to detect. 

\subsection{$\rho$-robustness}
Certified robustness establishes a threshold on the sizes of perturbations applied to neural network inputs. Perturbations smaller than the threshold do not affect the outputs of the classifier. Let $f_y(\cdot)$ denote the probability the output of the GNN associates the input ``$\cdot$'' to class $y \in \mathcal{Y}$. Below are the formal definitions used to describe attacks on neural networks and the robustness of neural networks to the attacks.

\textit{$\ell_p$-norm $\alpha$-bounded attack}:		For a given classifier $f(\cdot)$ and a fixed input $\mathbf{s}$, a successful \textit{$\ell_p$-norm $\alpha$-bounded attack} with $0 < \alpha < 1$ is one where the attacker can craft a perturbation $\vec{\xi}$ satisfying 
\begin{equation}
    \| \vec{\xi} \|_p \equiv \bigg( \sum_{n=1}^{|\vec{\xi}|} |\vec{\xi}_n|^p \bigg)^{\frac{1}{p}} \le |\vec{\xi}|\alpha,
\end{equation}
where $|\vec{\xi}|$ is the cardinality of $\vec{\xi}$, such that 
\begin{equation}
	\underset{y \in \mathcal{Y}}{\arg \max} ~  f_y(\mathbf{s} + \vec{\xi}) \ne \underset{y \in \mathcal{Y}}{\arg \max} ~ f_y(\mathbf{s}).
\end{equation}
	
\textit{$\ell_p$-norm $\rho$-robustness}: A classifier $f$ is \textit{$\ell_p$-norm $\rho$-robust} for fixed input $\mathbf{s}$ and $0 < \rho < 1$ if 
\begin{equation}
	\underset{y \in \mathcal{Y}}{\arg \max} ~  f_y(\mathbf{s} + \vec{\xi}) = \underset{y \in \mathcal{Y}}{\arg \max} ~ f_y(\mathbf{s})
\end{equation}
for all $\vec{\xi}$ satisfying $\|\vec{\xi}\|_p \le |\vec{\xi}| \rho$.
	
In our setting with graph data, the natural choice for $p$ is 1, i.e., the norm is the $\ell_1$-norm, and the perturbation uses the XOR operator $\oplus$ 
to append the attack vector.
	
\textit{Randomized smoothing (RS)} \cite{cohen2019certified,alfarra2022data} provides a certificate of robustness by constructing a smoothed classifier $g$ based on the base classifier $f$ as follows:
\begin{equation}
	g(\mathbf{s}) = \int \dd w_{\mathcal{N}(0,\sigma^2\mathbf{I})}(\vec{\zeta}) f(\mathbf{s}+\vec{\zeta}) = \langle f(\mathbf{s}+\vec{\zeta}) \rangle_{w_{\mathcal{N}(0,\sigma^2\mathbf{I})}},
\end{equation}
where $w_{\mathcal{N}(0,\sigma^2\mathbf{I})}(\cdot)$ is the probability density function (PDF) of a zero-mean Gaussian random vector with covariance matrix $\sigma^2\mathbf{I}$ and $\langle \cdot \rangle_w$ denotes mathematical expectation with respect to the PDF $w$. We will omit the subscript $w$ where the underlying probability law is understood from context.  
Suppose $y^*$ is the class with the highest probability predicted by $g$ for input $\mathbf{s}$, i.e., 
\begin{equation}
	y^* \equiv \arg \max_{y \in \mathcal{Y}} ~ g_y(\mathbf{s}).
\end{equation}
Define the probabilities
\begin{equation}
	p^* \equiv \max_{y \in \mathcal{Y}} ~ g_y(\mathbf{s}) = g_{y^*}(\mathbf{s})
\end{equation} 
and
\begin{equation}
	\widetilde{p} \equiv \max_{y \in \mathcal{Y} \setminus y^*} ~ g_y(\mathbf{s}),
\end{equation} 
where $p^* \ge \widetilde{p}$.  RS provides the following robustness certificate, which depends on input $\mathbf{s}$:
	\begin{equation}
		\underset{y \in \mathcal{Y}}{\arg \max} ~  g_y(\mathbf{s} + \xi) = \underset{y \in \mathcal{Y}}{\arg \max} ~ g_y(\mathbf{s})
	\end{equation}
	for all perturbations $\xi$ that satisfy $\|\xi\|_2 \le \rho$ with certification radius 
	\begin{equation}
		\rho = \frac{\sigma}{2} \big( \Phi^{-1}(p^*) - \Phi^{-1}(\widetilde{p}) \big),
	\end{equation}
	where $\Phi(\cdot)$ is the cumulative distribution function (CDF) of a standard Gaussian random variable, and $\sigma$ is the standard deviation parameter used to construct $g(\cdot)$.
 
\begin{mdframed}[style = myStyle]
\smallskip
Typically, $\rho$ is small. Increasing $\rho$ increases the risk of misclassification, albeit robustly, due to smoothing of the classification boundaries in the input space. Our approach complements $\rho$-robustness via input corrections, allowing for small values of $\rho$, mitigating misclassification due to smoothing.

\smallskip
\end{mdframed}

\section{The method}\label{sec:method}
The approach assumes that the GNN is $\ell_1$-norm $\rho$-robust. We aim to ensure that the distance between the corrected and true signals is within $\rho$ in the $\ell_1$ sense for any (user-selected) $\rho$. The approach can be envisaged as an extension to RS that overcomes the limited certification of RS.

\subsection{The encoding scheme (sender)}
The approach is a special kind of repetition code from information theory.  Let $K$ denote the number of repetitions of the original input $\mathbf{s}$, $\mathbf{1}_K$ is a vector of 1's of size $K$, and $\mathbf{t} \in \{0,1\}^{NK}$ denotes the encoded signal:
\begin{equation}
	\mathbf{t} = \mathbf{s} \otimes \mathbf{1}_K,
	\label{rep_1}
\end{equation}
where $\otimes$ is the Kronecker product.  The vector $\mathbf{t}$ consists of $K$ blocks of size $N$ of equal values: 
\begin{equation*}
	\mathbf{s} = \begin{bmatrix} s_1 \\ \vdots \\ s_N \end{bmatrix}  \mapsto  \mathbf{t} = \begin{bmatrix} \begin{bmatrix} s_{1,1} \\ \vdots \\ s_{1,K} \end{bmatrix} \\ \vdots \\ \begin{bmatrix} s_{N,1} \\ \vdots \\ s_{N,K} \end{bmatrix} \end{bmatrix},
	\label{rep_2}
\end{equation*}
where $s_{n,k} = s_{n,k'}$ for all $1 \le k,k'\le K$.  Essentially, each element $s_n$ of $\mathbf{s}$ is encoded into a $K$-vector of elements $s_{n,k} = s_n$ for all $1\le k \le K$.  

To obscure this repetitive structure from the attacker, the input signal is perturbed by a random vector, $\vec{\zeta} \in \{0,1\}^{NK}$, consisting of independent and identically distributed (i.i.d.) Bernoulli random variables, each with perturbation probability $\nu \equiv \PP(\zeta_i = 1)$ for all $1 \le i \le NK$.  The parameter $\nu$ is known to the defender but not the attacker.  

The attacker sees a series of $K$ defender-perturbed signals $(\mathbf{t} \oplus \vec{\zeta})|_1,...,(\mathbf{t} \oplus \vec{\zeta})|_K$, where
\begin{equation*}
	(\mathbf{t} \oplus \vec{\zeta})|_k = \mathbf{s} \oplus \vec{\zeta}|_k = \begin{bmatrix} s_1 \oplus \zeta_{k,1} \\ \vdots \\ s_N \oplus \zeta_{k,N}	\end{bmatrix}.
\end{equation*}  
The adversarial perturbation produces the vector 
\begin{equation}
	\mathbf{r} = (\mathbf{t} \oplus \vec{\zeta}) \oplus \vec{\xi},
\end{equation}
where $\vec{\xi}$ is an $NK$-dimensional vector of i.i.d. Bernoulli random variables with parameter $\beta$: from the perspective of the defender, any of the edges could be attacked with some unknown probability ($\beta$). The vector $\mathbf{r}$ consists of $N$ blocks of size $K$: 
\begin{equation*}
	\mathbf{r} = 
	\begin{bmatrix} \begin{bmatrix} s_{1,1}\oplus \zeta_{1,1}\oplus\xi_{1,1} \\ \vdots \\ s_{1,K} \oplus \zeta_{K,1} \oplus\xi_{K,1} \end{bmatrix} \\ \vdots \\ \begin{bmatrix} s_{N,1} \oplus \zeta_{N,1} \oplus\xi_{N,1} \\ \vdots \\ s_{N,K} \oplus \zeta_{N,K} \oplus\xi_{N,K} \end{bmatrix} \end{bmatrix}=\begin{bmatrix} \begin{bmatrix} r_{1,1} \\ \vdots \\ r_{1,K} \end{bmatrix} \\ \vdots \\ \begin{bmatrix} r_{N,1} \\ \vdots \\ r_{N,K} \end{bmatrix} \end{bmatrix}.
\end{equation*}

The perturbed version of element $s_n$ of $\mathbf{s}$ is associated with the $n$th block of $\mathbf{r}$.  We write $t_{n,k}$ and $r_{n,k}$ when referencing elements of $\mathbf{t}$ and $\mathbf{s}$ to emphasize the block structure of the code.

Our encoding scheme assumes that $\nu$, $N$, and $K$ are known to the defender but hidden from the attacker. The perturbation probability $\beta$ is not known to the defender. The attacker can see the size of the perturbed signal $NK$ but not the factors $N$ and $K$. The time complexity of the encoding scheme is $O(NK)$ since generating a Bernoulli random variate has the complexity of generating a uniform pseudo-random number on $(0,1)$, which is $O(1)$. As $N \propto |\mathcal{V}|^2$, the time complexity of the encoding scheme scales with the graph size as $O(K|\mathcal{V}|^2)$. However, the encoding scheme can be parallelized. One can achieve a best-case time complexity of $O(K)$ by encoding each of the $N$ blocks in parallel.

\subsection{The decoding scheme (receiver)}
This paper analyzes the following \emph{majority voting} scheme, which the receiver uses to decode the perturbed signal $\mathbf{r} = [r_{1,1} \cdots r_{1,K} \cdots r_{N,1} \cdots r_{N,K}]^{\top}$.  For given $K$ (known to the receiver), the receiver calculates the decoded signal, denoted $\widehat{\mathbf{s}}_K = [\widehat{s}_1~ \cdots~\widehat{s}_N]^{\top}$, as follows:
\begin{equation}
	\widehat{s}_n = \begin{cases}
		0 & \mbox{if } \frac{1}{K} \sum_{k=1}^K r_{n,k} \le 0.5 \\
		1 & \mbox{if } \frac{1}{K} \sum_{k=1}^K r_{n,k} > 0.5
	\end{cases} \mbox{ for } 1 \le n \le N. \label{eq_majority}
\end{equation}
That is, the receiver performs the mapping
\begin{equation*}
	\mathbf{r} = \begin{bmatrix} \begin{bmatrix} r_{1,1} \\ \vdots \\ r_{1,K} \end{bmatrix} \\ \vdots \\ \begin{bmatrix} r_{N,1} \\ \vdots \\ r_{N,K} \end{bmatrix} \end{bmatrix} \overset{\eqref{eq_majority}}{\longmapsto} \begin{bmatrix} \widehat{s}_1 \\ \vdots \\ \widehat{s}_N \end{bmatrix} = \widehat{\mathbf{s}}_K.
\end{equation*}

Since the entries of $\mathbf{r}$ are binary, the worst-case time complexity of the majority vote for each $n$ is $O(K)$. Thus, like the encoding scheme, decoding scales with the graph size as $O(K|\mathcal{V}|^2)$. However, unlike encoding, decoding cannot be parallelized.

\section{Analysis} \label{sec:analysis}
\subsection{Empirical probability of successful decoding}
For each $1 \le n \le N$ and $1 \le k \le K$, as $\vec{\xi}$ and $\vec{\zeta}$ are independent,
\begin{align}
	& \PP(\xi_{n,k} \oplus \zeta_{n,k} = 1) \nonumber \\& \quad = \PP(\xi_{n,k} = 1) \PP(\zeta_{n,k} = 0) + \PP(\xi_{n,k} = 0 ) \PP(\zeta_{n,k} = 1) \nonumber \\& \quad= \beta(1-\nu) + (1-\beta)\nu.
	\label{nu}
\end{align}
$\xi_{n,k} \oplus \zeta_{n,k} = 1$ represents the event that the (combined) perturbation changes the status of element $(n,k)$ of $\mathbf{t}$ (from 1 to 0 or 0 to 1).  The probability $\PP(\xi_{n,k} \oplus \zeta_{n,k} = 1)$ is a random variable as it depends on $\beta$, which is unknown to the defender. 

To simplify notation, we define
\begin{equation}
	\mu \equiv \PP(\xi_{n,k} \oplus \zeta_{n,k} = 1) = \beta(1-\nu) + (1-\beta)\nu.
\end{equation}
As $\beta < 0.5$  and $\nu < 0.5$, $\mu \in [\nu, 0.5)$. The probability of a successful recovery (based on a majority vote) is defined as:
\begin{equation}
	p_K = \frac{1}{N} \sum_{n=1}^N \PP \Big( \sum_{k=1}^K \mathbbm{1}_{\{1\}}(\xi_{n,k} \oplus \zeta_{n,k}) \le \frac{K}{2}-1 \Big), \label{eq_pk}
\end{equation}
that is, that \emph{no more than half} of the sample in each of the $N$ blocks is perturbed on average ($\mathbbm{1}_{\mathcal{X}}(x) = 1$ if $x \in \mathcal{X}$ and 0, otherwise). Without loss of generality, we will assume that $K$ is even. In a majority voting scheme, for each $1 \le n \le N$
\begin{equation}
	\sum_{k=1}^K \mathbbm{1}_{\{1\}}(\xi_{n,k} \oplus \zeta_{n,k}) \le \frac{K}{2}-1 ~ \Longleftrightarrow ~ \widehat{s}_n = s_n.
\end{equation}
(Here, `$\Leftrightarrow$' means \emph{if and only if}.) 
Consequently,
\begin{equation}
	p_K \ge 1-\rho ~ \Longleftrightarrow ~ \frac{1}{N} \| \widehat{\mathbf{s}}_K - \mathbf{s} \|_1 \le \rho. \label{eq_rho}
\end{equation}
In terms of $\mu$, $p_K$ is given by
\begin{align}
	p_K &= \frac{1}{N}\sum_{n=1}^N \sum_{k=0}^{\frac{K}{2}-1} \binom{K}{k} \mu^k ( 1 - \mu )^{K-k} \nonumber 
	\\ &= \sum_{k=0}^{\frac{K}{2}-1} \binom{K}{k} \mu^k ( 1 - \mu )^{K-k}. \label{eq_p_K}
\end{align}

The majority vote scheme \eqref{eq_majority} informs a partitioning of the set $\{1,...,N\}$ into two subsets based on the result of the majority vote:
\begin{equation}
	\mathcal{N}_0 \equiv \Big\{n: \frac{1}{K} \sum_{k=1}^K r_{n,k} \le 0.5 \Big\}
\end{equation}
and
\begin{equation}
	\mathcal{N}_1 \equiv \Big\{n: \frac{1}{K} \sum_{k=1}^K r_{n,k} > 0.5 \Big\}.
\end{equation}
To simplify notation, we will define
\begin{equation}
	R_{n,K} \equiv \frac{1}{K} \sum_{k=1}^K r_{n,k}.
\end{equation}

The probability $\mu$ is unknown to the defender; from their perspective, it is a random variable.  
We consider the following estimator of $p_K$:
\begin{multline}
	\widehat{p}_K= \frac{1}{N}\sum_{n=1}^N \sum_{k=0}^{\frac{K}{2}-1} \binom{K}{k} \Big({R_{n,K_1}}^k ( 1 - R_{n,K_2} )^{K-k} \mathbbm{1}_{\mathcal{N}_0}(n)
	\\+ (1 - R_{n,K_1})^k {R_{n,K_2} }^{K-k} \mathbbm{1}_{\mathcal{N}_1}(n) \Big),
	\label{eq_p_Khat}
\end{multline}
where
\begin{equation}
	R_{n,K_1} = \frac{2}{K} \sum_{k = 1}^{\frac{K}{2}} r_{n,k}
\end{equation}
and
\begin{equation}
	R_{n,K_2} = \frac{2}{K} \sum_{k = \frac{K}{2}+1}^{K} r_{n,k}.
\end{equation}

Essentially, $R_{n,K_1}$ and $R_{n,K_2}$ are two independent but identically distributed means taken over two halves of the sample $\{r_{n,1},...,r_{n,K}\}$.  Moreover,
\begin{equation}
	R_{n,K} = \frac{1}{2} \big( R_{n,K_1} + R_{n,K_2} \big).
\end{equation} 
We show in the next section that this splitting of the sample results in $\widehat{p}_K$ to be an unbiased estimator of $p_K$.

The empirical probability of successful recovery, $\widehat{p}_K $, approximates the true probability with increasing accuracy as $K$ gets larger. In the next section, we show that the difference can be made arbitrarily small with an appropriate choice of $K$.

\subsection{Selection of $K$ and performance guarantee}
\begin{mdframed}[style = myStyle]

\smallskip

\textsf{The main result}: 
Suppose the GNN is $\ell_1$-norm $\rho$-robust and assume that the perturbation probability $\mu = \PP(\xi_{n,k} \oplus \zeta_{n,k} = 1) < 0.5$. Then, for any $0 < \eta \ll 1$ and any $0 < \rho \ll 1$, there exists $K$ so that
\begin{equation}
	\PP\Big( \frac{1}{N} \| \widehat{\mathbf{s}}_K - \mathbf{s} \|_1 \le \rho \Big) \ge 1- \eta.
\end{equation}

To achieve this, we select $K$ so that
\begin{align}
	\mbox{(i) } & \PP(|\widehat{p}_K - p_K| \ge \epsilon_{\mathrm{tol}}) \le \eta \mbox{ and} \label{eq_i}\\
	\mbox{(ii) } & \widehat{p}_K \ge 1 +  \epsilon_{\mathrm{tol}} - \rho, \label{eq_ii}
\end{align}
where $0 < \epsilon_{\mathrm{tol}} < \rho$ is pre-defined error threshold. Assuming condition (i) holds, condition (ii) simply ensures that $p_K \ge 1 - \rho$, which implies $\rho$-robustness; see  \eqref{eq_rho}.

\smallskip

\end{mdframed}

We start with the required $K$ to satisfy condition (i). For each $1 \le n \le N$, we have that
\begin{equation}
	R_{n,K} = 
	\begin{cases}
		\frac{1}{K} \sum_{k = 1}^K \xi_{n,k} \oplus \zeta_{n,k} & \mbox{if } t_{n,k} = s_n = 0 \\
		1 - \frac{1}{K} \sum_{k = 1}^K \xi_{n,k} \oplus \zeta_{n,k} & \mbox{if } t_{n,k} = s_n = 1
	\end{cases}. \label{eq_rnk1}
\end{equation}
From \eqref{eq_rnk1} we have that
\begin{multline}
	\PP(R_{n,K} \le 0.5) \\=  
	\begin{cases}
		\PP \Big(\sum_{k = 1}^K \xi_{n,k} \oplus \zeta_{n,k} \le \frac{K}{2} \Big) & \mbox{if } t_{n,k} = s_n = 0 
		\\ \PP \Big(\sum_{k = 1}^K \xi_{n,k} \oplus \zeta_{n,k} > \frac{K}{2} \Big) & \mbox{if } t_{n,k} = s_n = 1
	\end{cases}.
\end{multline}

For $n \in \mathcal{N}_0$ and all $1 \le k \le K$, $\langle r_{n,k} \rangle = \langle \xi_{n,k} \oplus \zeta_{n,k} \rangle = \mu$. Similarly, for $n \in \mathcal{N}_1$, $\langle r_{n,k} \rangle = 1 - \langle \xi_{n,k} \oplus \zeta_{n,k} \rangle =  1 - \mu$. This can also be used to determine higher moments of $r_{n,k}$: For $n \in \mathcal{N}_0$, $\langle (r_{n,k})^j \rangle =   \PP (\xi_{n,k} \oplus \zeta_{n,k}  = 1) = \mu$ and for $n \in \mathcal{N}_1$, $\langle (r_{n,k})^j \rangle =   \PP (\xi_{n,k} \oplus \zeta_{n,k} = 0) = 1-\mu$, for any $j$. Moreover, for $n \in \mathcal{N}_0$
\begin{align} 
		 &\big\langle {R_{n,K_1}}^j \big\rangle = \frac{2^j}{K^j} \Big\langle \Big( \sum_{k = 1}^{\frac{K}{2}} r_{n,k} \Big)^j \Big\rangle \nonumber
		\\& \quad = \frac{2^j}{K^j} \sum_{ \substack{\{j_k \ge 0\}: \\j_1 + \cdots + j_{K/2} = j } } \binom{j}{j_1, \cdots, j_{K/2} } \prod_{k=1}^{\frac{K}{2}} \big\langle (r_{n,k})^{j_k}  \big\rangle \nonumber
		\\& \quad = \frac{2^j}{K^j} \sum_{ \substack{\{j_k \ge 0\}: \\j_1 + \cdots + j_{K/2} = j } } \binom{j}{j_1, \cdots, j_{K/2} } \prod_{k=1}^{\frac{K}{2}} \mu 
		= \mu^j \label{eq_exp}
\end{align} 
and
\begin{equation} 
	\big\langle (1-R_{n,K_2})^j \big\rangle = (1 - \mu)^j.
\end{equation} 
Similarly, for $n \in \mathcal{N}_1$, 
\begin{equation}
	\big\langle (1 - R_{n,K_1})^j \big\rangle = \mu^j 
\end{equation}
and
\begin{equation}
	\big\langle {R_{n,K_2}}^j \big\rangle = (1 - \mu)^j.
\end{equation}

We use the moments above to show that $\widehat{p}_K$ is an unbiased estimator of $p_K$:
\begin{align}
	& \big\langle \widehat{p}_K \big\rangle  \nonumber \\ 
	& ~ = \frac{1}{N}\sum_{n=1}^N \sum_{k=0}^{\frac{K}{2}-1} \binom{K}{k} \Big( \mathbbm{1}_{\mathcal{N}_0}(n) \big\langle {R_{n,K_1}}^k (1-R_{n,K_2})^{K-k} \big\rangle \nonumber 
	\\& \qquad \qquad \qquad \quad + \mathbbm{1}_{\mathcal{N}_1}(n) \big\langle (1-R_{n,K_1})^k {R_{n,K_2}}^{K-k} \big\rangle \Big) \nonumber \\
	& ~ =  \frac{1}{N} \sum_{n=1}^N \sum_{k=0}^{\frac{K}{2}-1} \binom{K}{k} \mu^k (1-\mu)^{K-k} = p_K.
\end{align}
Thus, $\langle |\widehat{p}_K - p_K| \rangle = 0$ for all $K > 1$ (and even).  As a result, we have by appeal to McDiarmid's inequality that
\begin{equation}
	\PP(|\widehat{p}_K - p_K|  \ge \epsilon_{\mathrm{tol}})
	\le 2 \exp \Big( -\frac{2 \epsilon_{\mathrm{tol}}^2 }{N K \Delta_K^2} \Big),
\end{equation}
where $\Delta_K$ is a finite difference constant that binds (from above) the change in $\widehat{p}_K$ as a result of changing one $r_{n,k}$  (from 0 to 1 or 1 to 0). We turn to estimating $\Delta_K$ and use the inequality above to find a bound on $K$.

The maximal change in $R_{n,K_1}$ and $R_{n,K_2}$ is bounded from above by $2/K$. We, thus, consider a change of this magnitude in $R_{n,K_1}$ for some $n \in  \mathcal{N}_0$. A change in $R_{n,K_2}$ or in $\mathcal{N}_1$ results in the same bound. The change in $\widehat{p}_K$ as a result of modifying one $r_{n,k}$ is
\begin{equation}
	\frac{1}{N} \sum_{k=0}^{\frac{K}{2}-1} \binom{K}{k} \Big[ \Big( R_{n,K_1} + \frac{2}{K} \Big)^k - {R_{n,K_1}}^k \Big] (1 - R_{n,K_2})^{K-k}.
\end{equation}

We start by seeking an upper bound on the difference inside the sum $$\Big[ \Big( R_{n,K_1} + \frac{2}{K} \Big)^k - {R_{n,K_1}}^k \Big].$$ We have that
\begin{align}
		&\Big( R_{n,K_1} + \frac{2}{K} \Big)^k - {R_{n,K_1}}^k = \sum_{m=0}^{k-1} \binom{k}{m} {R_{n,K_1}}^m \Big( \frac{2}{K} \Big)^{k-m} \nonumber \\
		& \qquad \le \frac{2}{K} \sum_{m=0}^{k-1} \Big(\frac{m^2}{k-1} + 1 \Big) \binom{k-1}{m} {R_{n,K_1}}^m \Big( \frac{2}{K} \Big)^{k-1-m} \nonumber \\
		& \qquad \le \frac{2k}{K}\Big(R_{n,K_1} + \frac{2}{K} \Big)^{k-1}.
\end{align}
Hence, for $n \in \mathcal{N}_0$, the maximal change resulting from modifying one $r_{n,k}$ in $R_{n,K_1}$ is
\begin{align}
		&\frac{2}{NK} \sum_{k=0}^{\frac{K}{2}-1} k \binom{K}{k} \Big(R_{n,K_1} + \frac{2}{K} \Big)^{k-1} (1 - R_{n,K_2})^{K-k} \nonumber \\
		& \quad \le \frac{4}{N(K+4)} \sum_{k=1}^{\frac{K}{2}-1} k \binom{K}{k} \Big(R_{n,K_1} + \frac{2}{K} \Big)^k (1 - R_{n,K_2})^{K-k} \nonumber \\
		& \quad \le \frac{8e^2}{NK(K+2)}.
\end{align}

We get the same bound when the change in $r_{n,k}$ modifies $R_{n,K_1}$ and, by symmetry, for $n \in \mathcal{N}_1$.  Hence,
\begin{equation}
	\Delta_K = \frac{8e^2}{NK(K+2)}.
\end{equation}

\begin{figure}[h]
\centering
\includegraphics[width=0.5\textwidth,trim={0 .5cm 0 .5cm},clip]{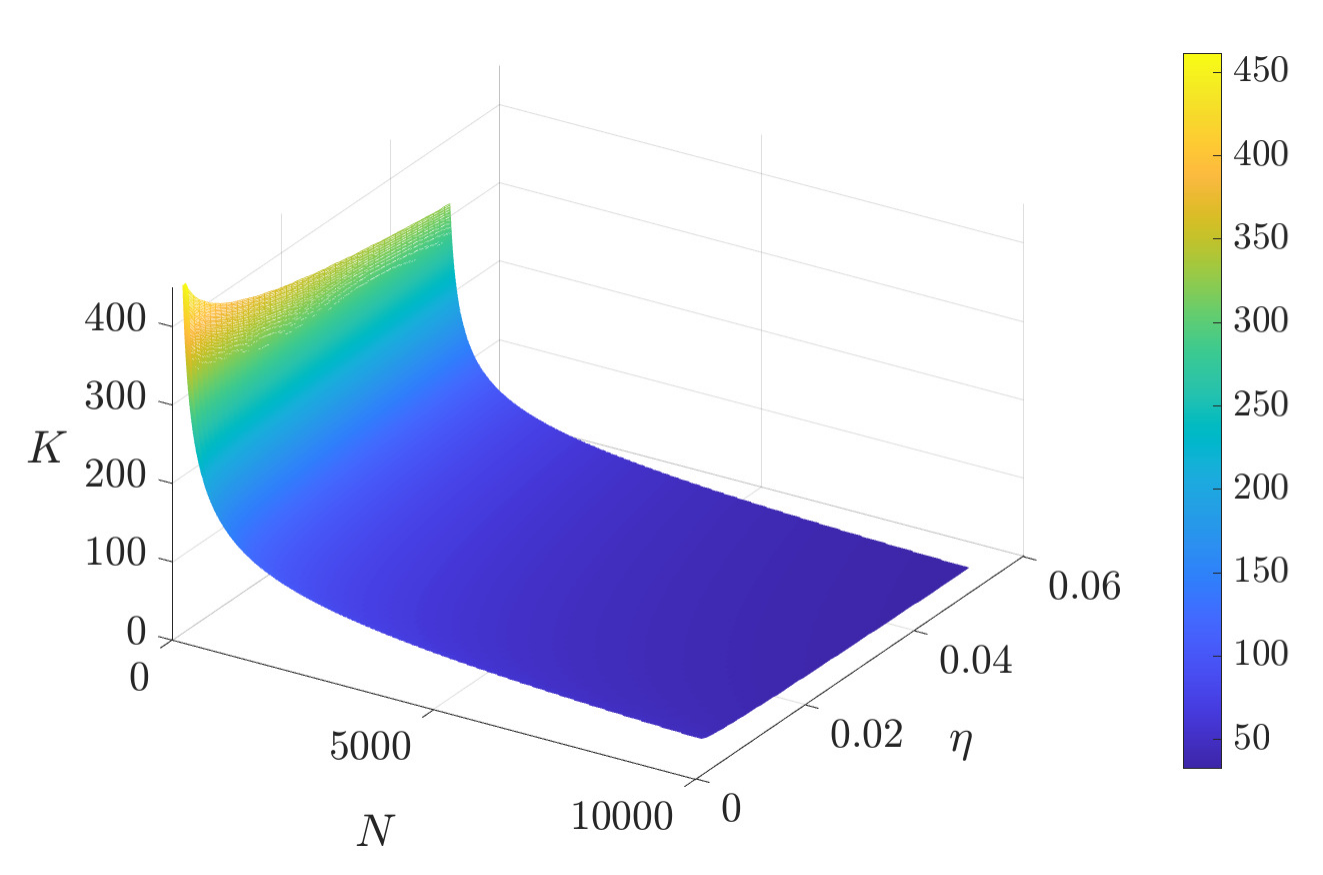}
\caption{$K$ versus $N$ and $\eta$ for $\epsilon_{\rm tol} = 0.025$.}
\label{fig:NvK}
\end{figure}

Thus, any $K$ that satisfies
\begin{equation}
	K^2 + 2K \ge \frac{-32 e^4}{\epsilon_{\mathrm{tol}}^2 N} \log_e \Big( \frac{\eta}{2} \Big) \label{eq_Ki}
\end{equation}
also satisfies condition (i) in \eqref{eq_i}. Thus, $K$ scales favorably with the size of the graph $N$ as illustrated in Fig.~\ref{fig:NvK}. 

This $K$ also ensures that 
\begin{equation}
	\PP(|\widehat{\mu} - \mu|  \ge \epsilon_{\mathrm{tol}})
	\le \eta,
\end{equation}
where 
\begin{equation}
	\widehat{\mu} = \frac{1}{N} \sum_{n \in \mathcal{N}_0} R_{n,K} + \frac{1}{N} \sum_{n \in \mathcal{N}_1} (1-R_{n,K})
\end{equation}
is an estimator of $\mu$. 
To satisfy condition (ii) in \eqref{eq_ii}, we start with $K$ based on \eqref{eq_Ki} and check if the condition is satisfied. If not, one can solve the following for $\widetilde{K}$ using a line search method.
\begin{equation}
	\sum_{k=0}^{\frac{\widetilde{K}}{2}-1} \binom{\widetilde{K}}{k} \widehat{\mu}^k ( 1 - \widehat{\mu} )^{\widetilde{K}-k} = 1 + \epsilon_{\mathrm{tol}} - \rho. \label{eq_Kii}
\end{equation}
and pick the smallest even $K \ge \widetilde{K}$. This will ensure that condition (ii) in \eqref{eq_ii} is also satisfied. 

\section{Experiments} \label{sec:experiments}
To test the method, we consider an attacker performing two types of perturbations to a randomly generated graph, i.e., the Erd\H{o}s-R\'{e}nyi model: the first is a random perturbation to the edges of the graph, each with probability 0.2, and the second is one where the attacker disconnects the vertex with the highest eigenvector centrality. Throughout our experiments, we treat $\rho =0.05$ as a reasonable robustness target; that is, the classifiers can handle errors of up to 5\% in the inputs. 

\newpage
\onecolumngrid
\begin{figure*}[!]
	\centering
	\subfloat[$N = 1000$]{\includegraphics[width=0.4\textwidth,trim={0 0cm 0 0.5cm},clip]{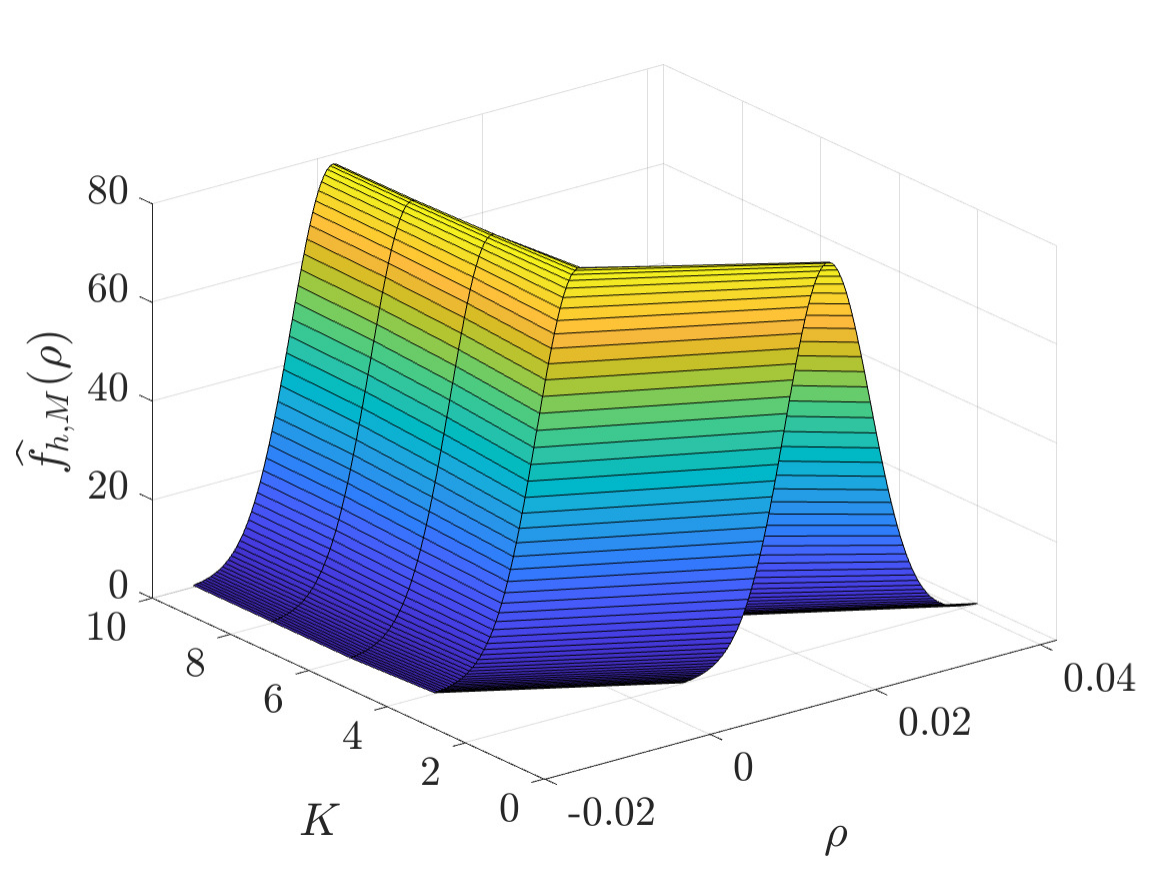}%
	}
	\subfloat[$N = 100$]{\includegraphics[width=0.4\textwidth,trim={0 0cm 0 0.5cm},clip]{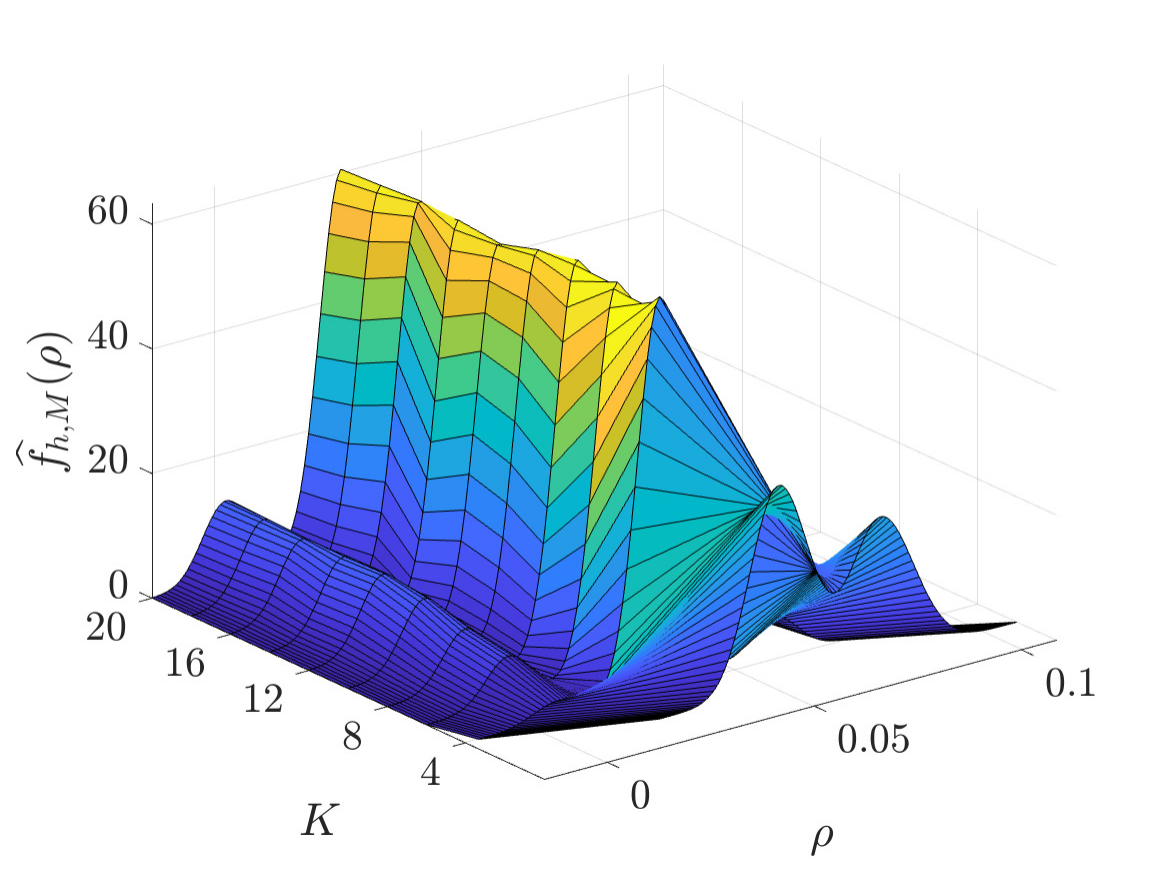}}
	
	\subfloat[$N = 50$]{\includegraphics[width=0.4\textwidth,trim={0 0cm 0 0.5cm},clip]{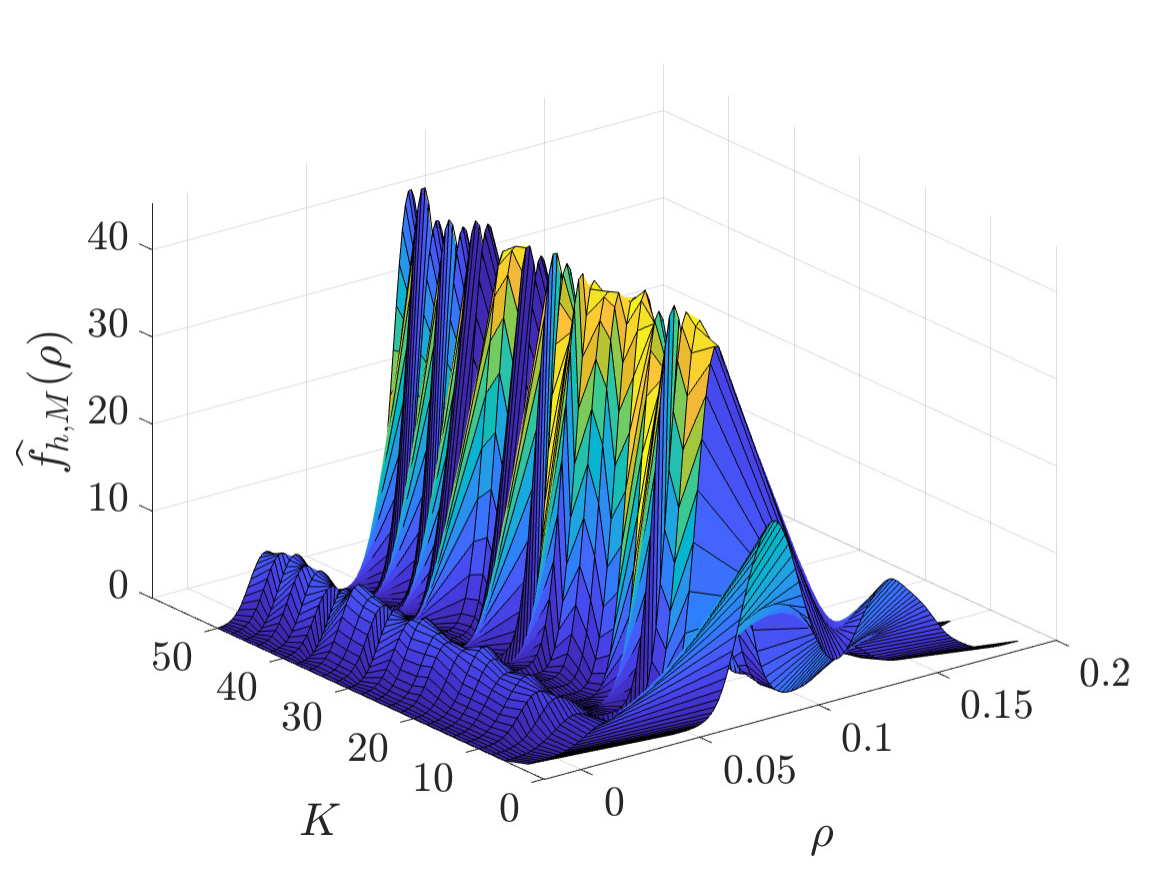}%
	}
	\caption{Empirical probability density functions of  $\frac{1}{N} \| \widehat{\mathbf{s}}_K - \mathbf{s} \|_1$ for different values of $K$, calculated using kernel density estimators, $\widehat{f}_{h,M}$, over $M=1000$ randomly simulated graphs, attacked, and reconstructed using the proposed approach. In all cases, $h = 0.005$ to avoid negative probabilities. We employed a mild sender-assigned perturbation of $\nu = 0.01$. For $N=1000$, (a), we observe that $\widehat{f}_{h,M}$ ceases to change after $K=4$, indicating that larger graphs require fewer repetitions to correct. For $N = 100$ and $N=50$, (b) and (c), respectively, we observe slight oscillatory behavior in $\widehat{f}_{h,M}$. The distributions stabilize after $K = 6$ in (b) and $K = 14$ in (c). The peaks in the stable distributions are approximately (a) $\rho_{\rm peak} \approx 0.003$, (b) $\rho_{\rm peak} \approx 0.03$, (c) $\rho_{\rm peak} \approx 0.07$.}
	\label{fig:fHat}
\end{figure*}
\twocolumngrid

We summarize the results using empirical probabilities, namely kernel density estimators (KDEs):
\begin{multline}
    \widehat{f}_{h,M}(\rho) = \frac{1}{hM} \sum_{m=1}^M w_{\mathcal{N}(N^{-1}\|\widehat{\mathbf{s}}_K^{[m]} - \mathbf{s}^{[m]} |_1,h)} (\rho) \\
    = \frac{1}{\sqrt{2\pi}hM} \sum_{m=1}^M  \exp \bigg( - \frac{1}{2h^2} \big(N^{-1}\big\|\widehat{\mathbf{s}}_K^{[m]} - \mathbf{s}^{[m]} \big\|_1 - \rho\big)^2 \bigg),
\end{multline}
where $h$ is the bandwidth parameter, $\widehat{\mathbf{s}}_K^{[m]}$ and $\mathbf{s}^{[m]}$ are the decoded and true graph vectors in the $m$th simulation, and $M$ is the number of simulations (in our experiments $M = 1000$). We also employ an empirical cumulative distribution function (ECDF)
\begin{equation*}
    \widehat{F}_M(\rho) \approx \PP\Big( \frac{1}{N} \| \widehat{\mathbf{s}}_K - \mathbf{s} \|_1 \le \rho \Big)
\end{equation*}
given by
\begin{equation}
    \widehat{F}_M(\rho) = \frac{1}{M} \sum_{m=1}^M H \big( \rho - N^{-1} \big\| \widehat{\mathbf{s}}_K^{[m]} - \mathbf{s}^{[m]} \big\|_1 \big),
\end{equation}
where $H$ is the Heaviside step function.

We generate $M = 1000$ random graphs, where we fix the number of vertices ($N = 1000$, $N = 100$, $N = 50$) and connect each pair of vertices randomly and with a predefined probability of 0.2.  

\begin{figure*}[!]
	\centering
	\subfloat[$N = 100$, $\nu = 0.01$]{\includegraphics[width=0.4\textwidth,trim={0 0cm 0 0.5cm},clip]{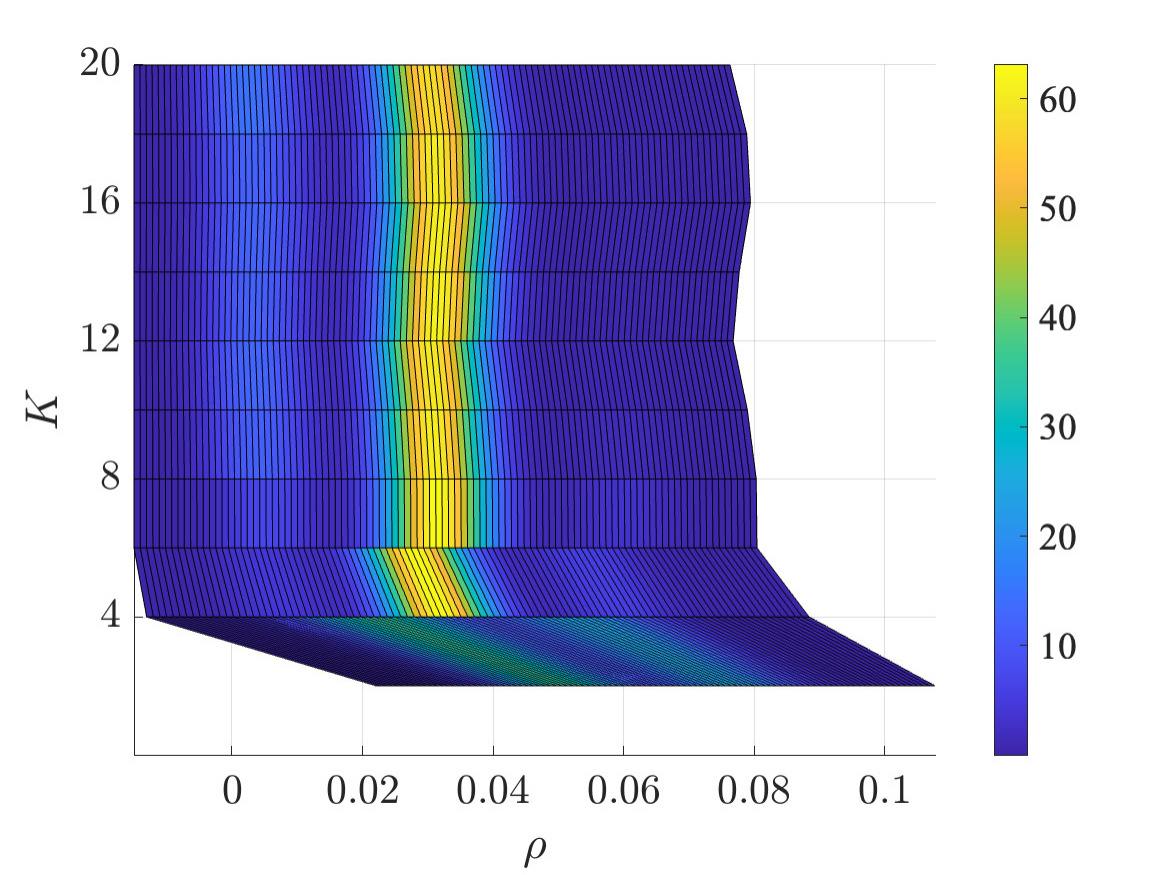}%
	}
	\subfloat[$N = 100$, $\nu = 0.1$]{\includegraphics[width=0.4\textwidth,trim={0 0cm 0 0.5cm},clip]{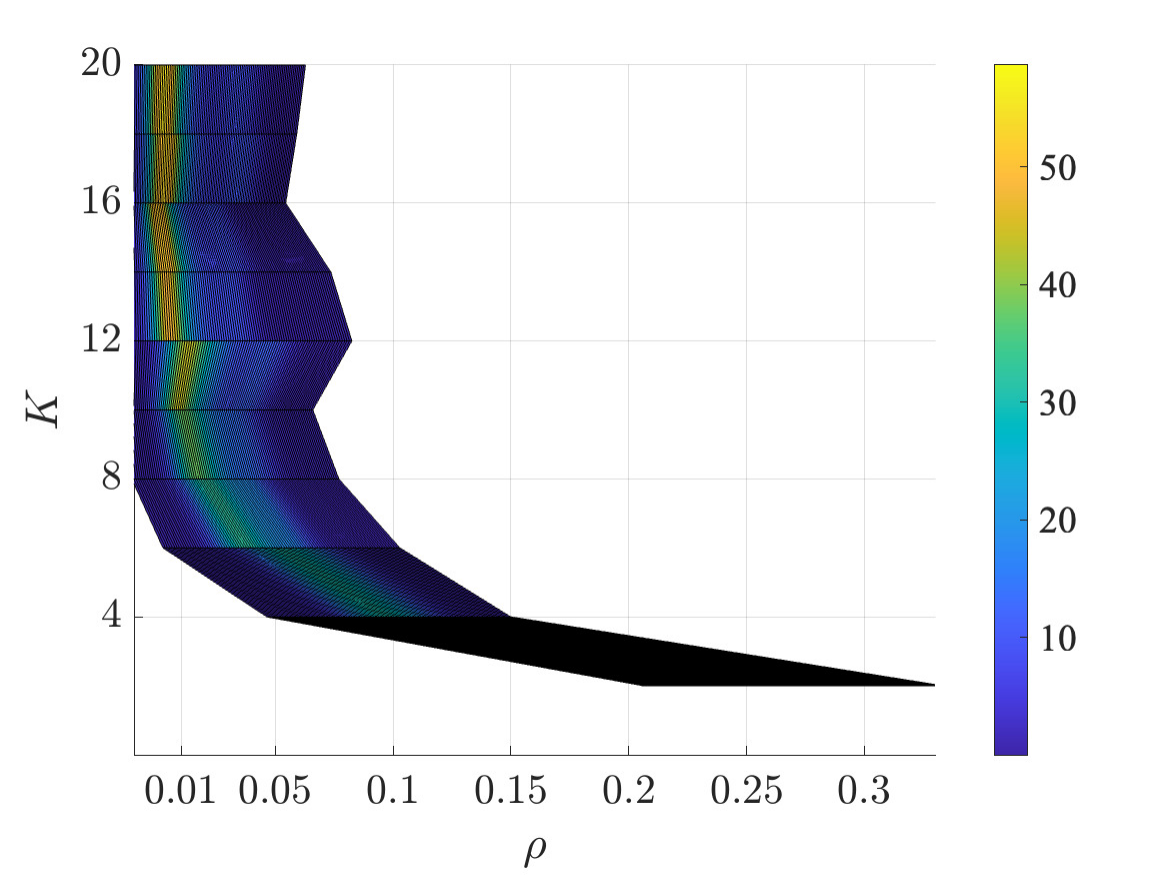}%
	}
	\caption{Top view of $\widehat{f}_{h,M}$ for $N=100$. (a) When $\nu = 0.01$, we see that $\widehat{f}_{h,M}$ stabilizes at $K=6$ with a peak around $\rho_{\rm peak} = 0.03$. (b) When $\nu = 0.1$, we see that $\widehat{f}_{h,M}$ stabilizes late (around $K=12$) but at a smaller peak around $\rho_{\rm peak} = 0.01$.}
	\label{fig:fHatVsNu}
\end{figure*}

Our first tests use a small sender-assigned noise probability of $\nu = 0.01$. The empirical probability densities, represented by $\widehat{f}_{h,M}$, are shown in Fig.~\ref{fig:fHat}. We make two observations: For larger graphs, e.g., $N=1000$, a smaller number of repetitions is required to achieve accurate correction. Second, both the number of repetitions needed and the location of the peak increase as the graph gets smaller. One can mitigate this by increasing $\nu$: see Fig.~\ref{fig:fHatVsNu}. 

\newpage
\onecolumngrid

\begin{figure*}[!h]
	\centering
	\subfloat[$N = 1000$]{\includegraphics[width=0.33\textwidth,trim={0 0cm 0 0.5cm},clip]{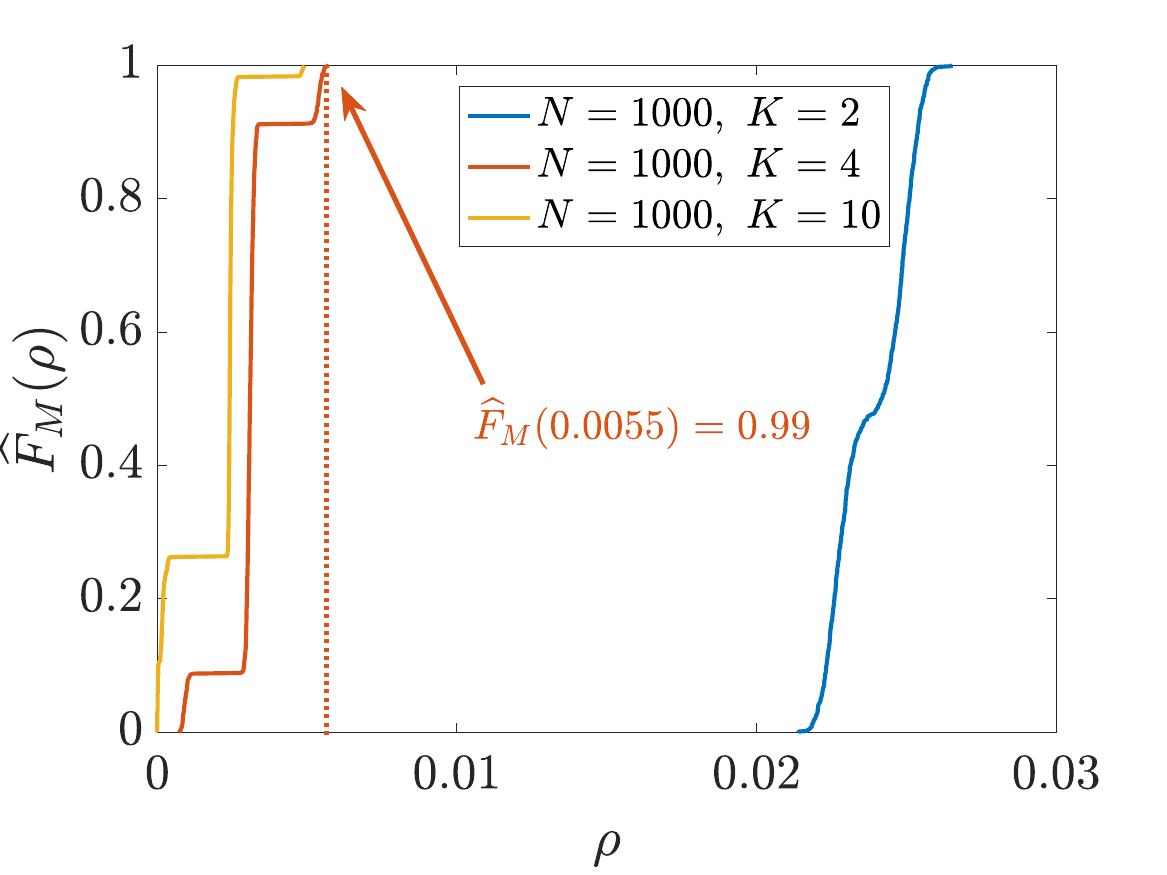}%
	}
	\subfloat[$N = 100$]{\includegraphics[width=0.33\textwidth,trim={0 0cm 0 0.5cm},clip]{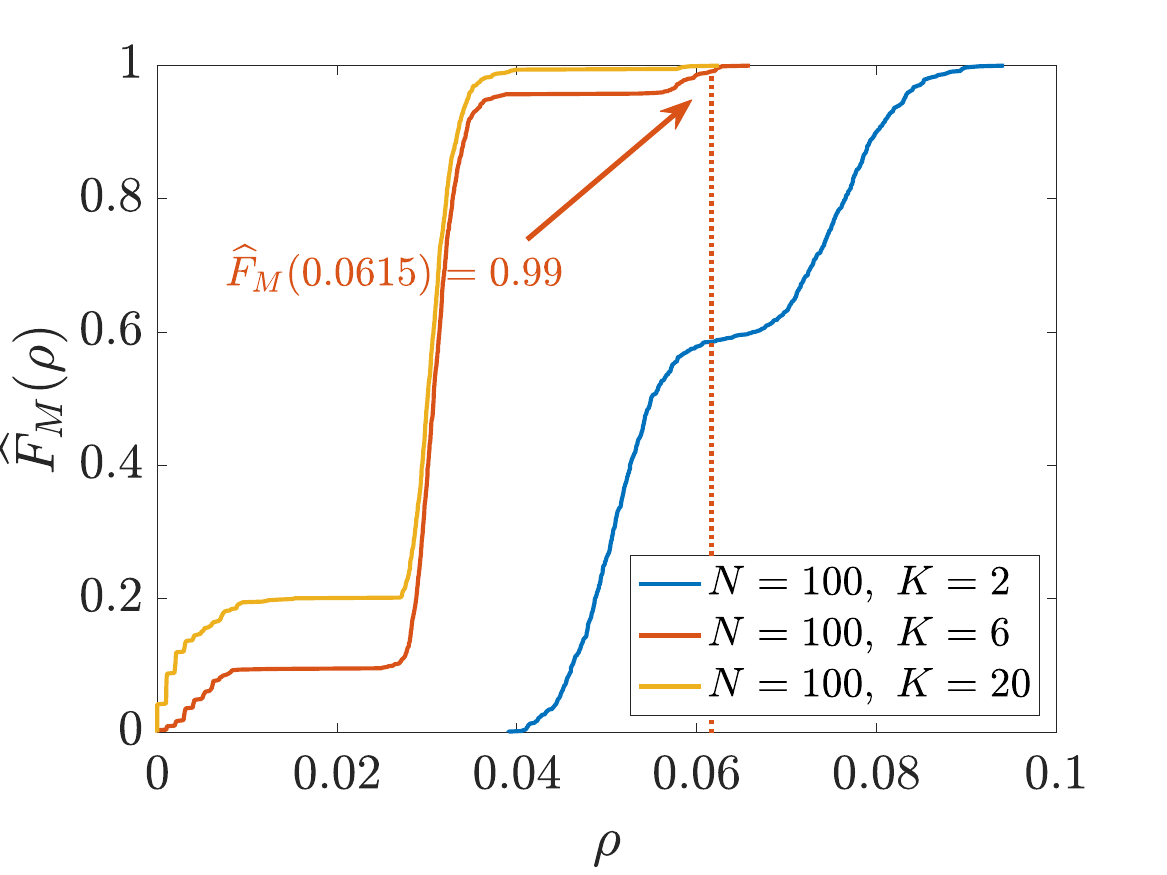}}
	\subfloat[$N = 50$]{\includegraphics[width=0.33\textwidth,trim={0 0cm 0 0.5cm},clip]{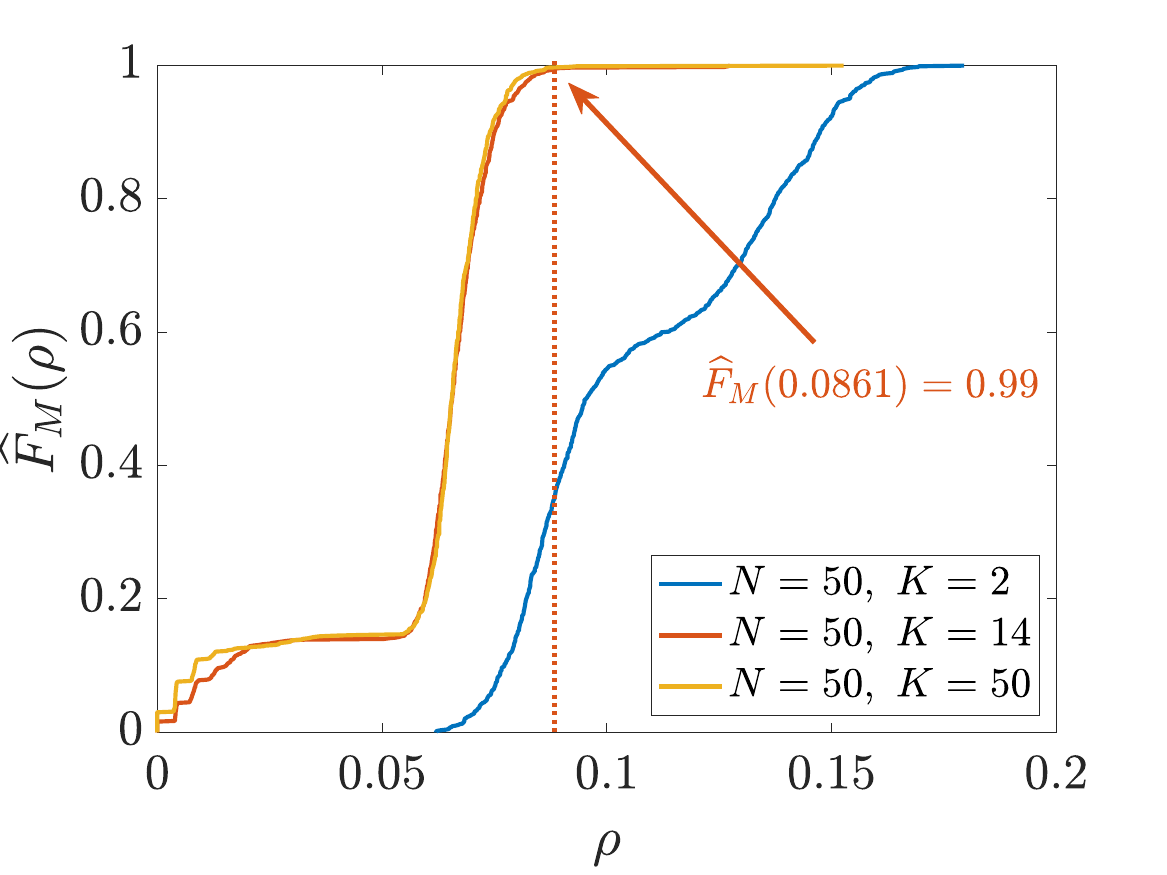}%
	}
	\caption{ECDFs of  $\frac{1}{N} \| \widehat{\mathbf{s}}_K - \mathbf{s} \|_1$ illustrating convergence to stable probability distributions at (a) $K = 4$ when $N = 1000$, (b) $K = 6$ when $N = 100$, and (c) $K = 14$ when $N = 50$. The settings are the same as those shown in Fig.~\ref{fig:fHat}, namely, $\nu = 0.01$. We see that (a) $\PP(\frac{1}{N} \| \widehat{\mathbf{s}}_4 - \mathbf{s} \|_1 \le 0.0055) \approx 0.99$ when $N=1000$, which is a very reasonable reconstruction: well within the bounds of $\rho$-robustness one expects from adversarial training in practice. For the smaller graphs, (b) $N=100$ and (c) $N=50$, we see that $\PP(\frac{1}{N} \| \widehat{\mathbf{s}}_6 - \mathbf{s} \|_1 \le 0.0615) \approx 0.99$ and $\PP(\frac{1}{N} \| \widehat{\mathbf{s}}_{14} - \mathbf{s} \|_1 \le 0.0861) \approx 0.99$, respectively. The $\rho$ values of 0.0615 and 0.0861 are somewhat large (greater than our acceptable threshold value of $\rho = 0.05$), indicating that the sender-assigned perturbation may need to be larger for smaller graphs.}
	\label{fig:FHat}
\end{figure*}
\twocolumngrid

We'll explore this further for small graphs below but let's first quantify  $\PP(\frac{1}{N} \| \widehat{\mathbf{s}}_K - \mathbf{s} \|_1 \le \rho)$ more precisely through the simulations and $\widehat{F}_M$.

\begin{figure}[!]
	\centering
	\includegraphics[width=0.33\textwidth,trim={0 0cm 0 0.5cm},clip]{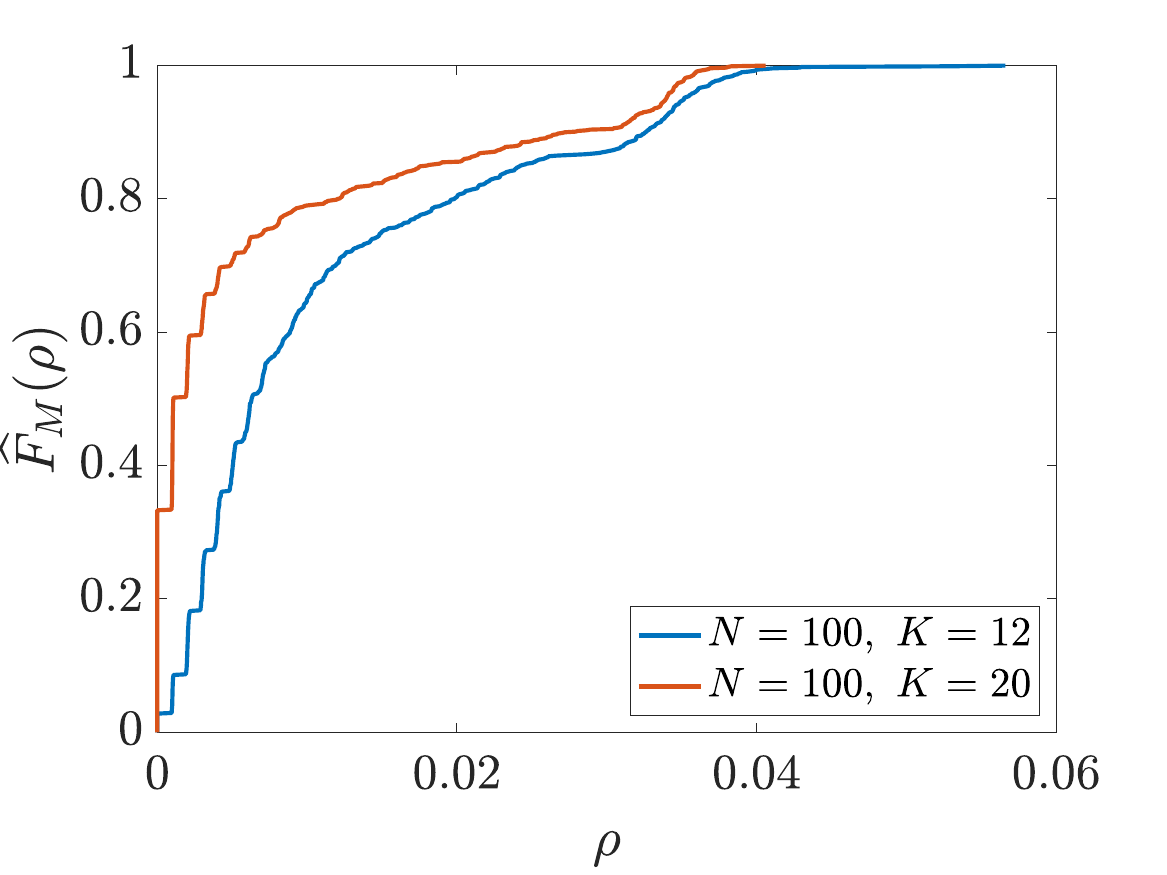}%
	\caption{$\widehat{F}_{M}$ for $N=100$ and $\nu = 0.1$; see Fig.~\ref{fig:fHatVsNu}b. We see perfect reconstruction with $K = 12$ for $\rho = 0.04$. In this case, our theory, \eqref{eq_Ki} would prescribe $K = 240$, much larger than required for accurate reconstruction.}
	\label{fig:FhatN100}
\end{figure}

Fig.~\ref{fig:FHat} shows ECDFs corresponding to the same experiments illustrated in Fig.~\ref{fig:fHat}. Convergence to stable distributions can be clearly seen in the figures. We also see that small sender-assigned perturbations are needed for larger graphs, namely, $N = 1000$. In this case, the non-random part of the attack is not pronounced, and our method recovers the graphs with values of $K$ that are significantly smaller than those prescribed by our bound \eqref{eq_Ki}; see Fig.~\ref{fig:FhatN100}.  We thus turn to smaller graphs to investigate the validity of our bounds.

\begin{figure}[!]
	\centering
	\subfloat[$N = 50$, $\nu = 0.1$]{\includegraphics[width=0.4\textwidth,trim={0 0cm 0 0.5cm},clip]{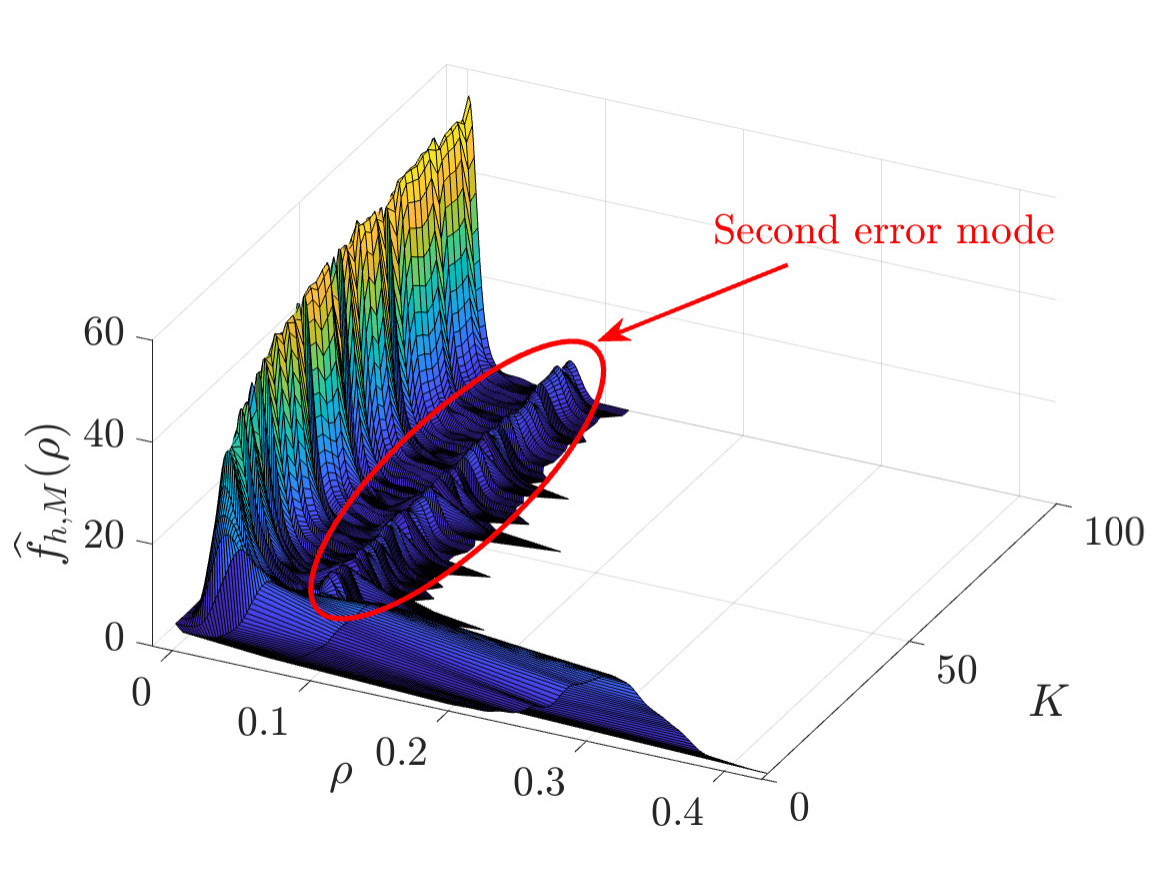}%
	}
	
	\subfloat[$N = 50$, $\nu = 0.25$]{\includegraphics[width=0.4\textwidth,trim={0 0cm 0 0.5cm},clip]{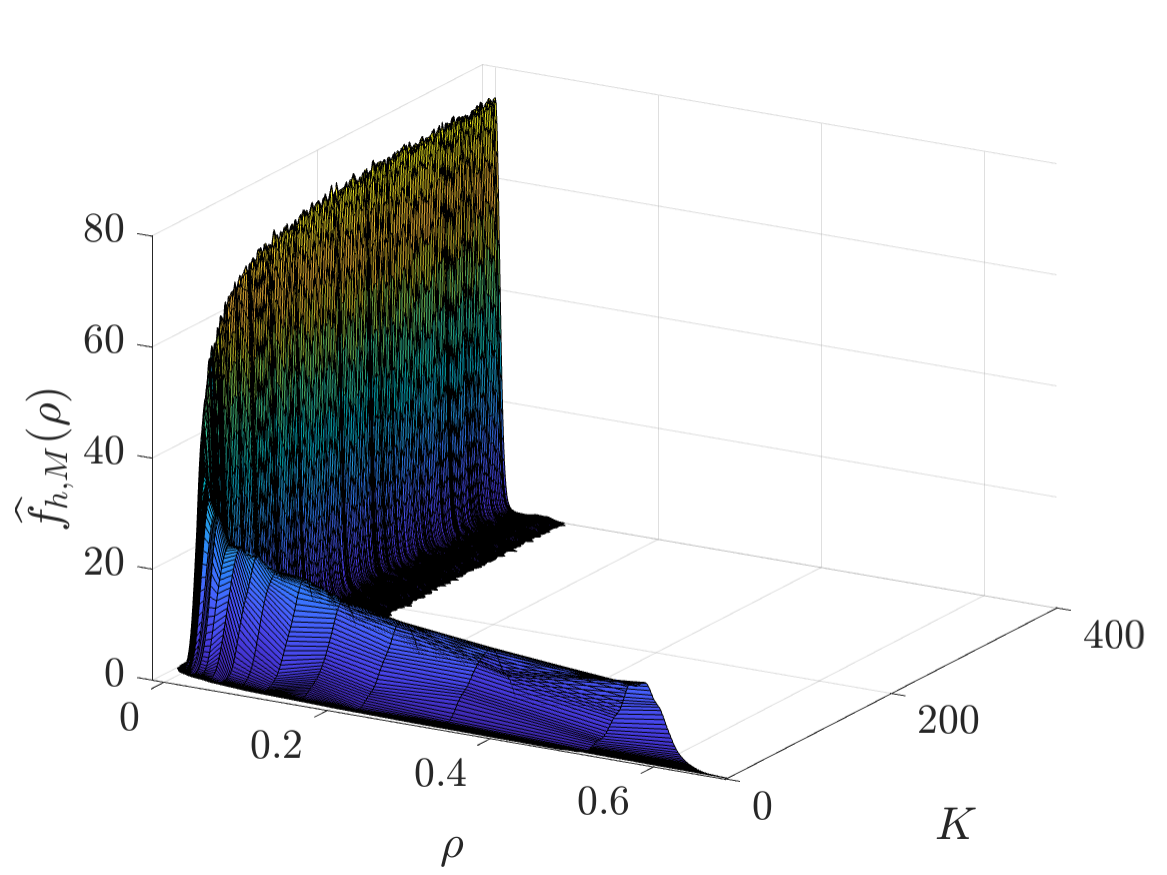}%
	}
	\caption{$\widehat{f}_{h,M}$ for $N=50$. (a) When $\nu = 0.1$, we see a second error mode persists across all values of $K$ around $\rho = 0.08$ similar to the case observed in Fig.~\ref{fig:FHat}c, indicating that this corresponds to edges removed by disconnecting the vertex with the highest eigenvector centrality. (b) Setting $\nu = 0.25$ consistently removes the second mode.}
	\label{fig:fHatVsNuN50}
\end{figure}

\begin{figure}[!]
	\centering
\includegraphics[width=0.33\textwidth,trim={0 0cm 0 0.5cm},clip]{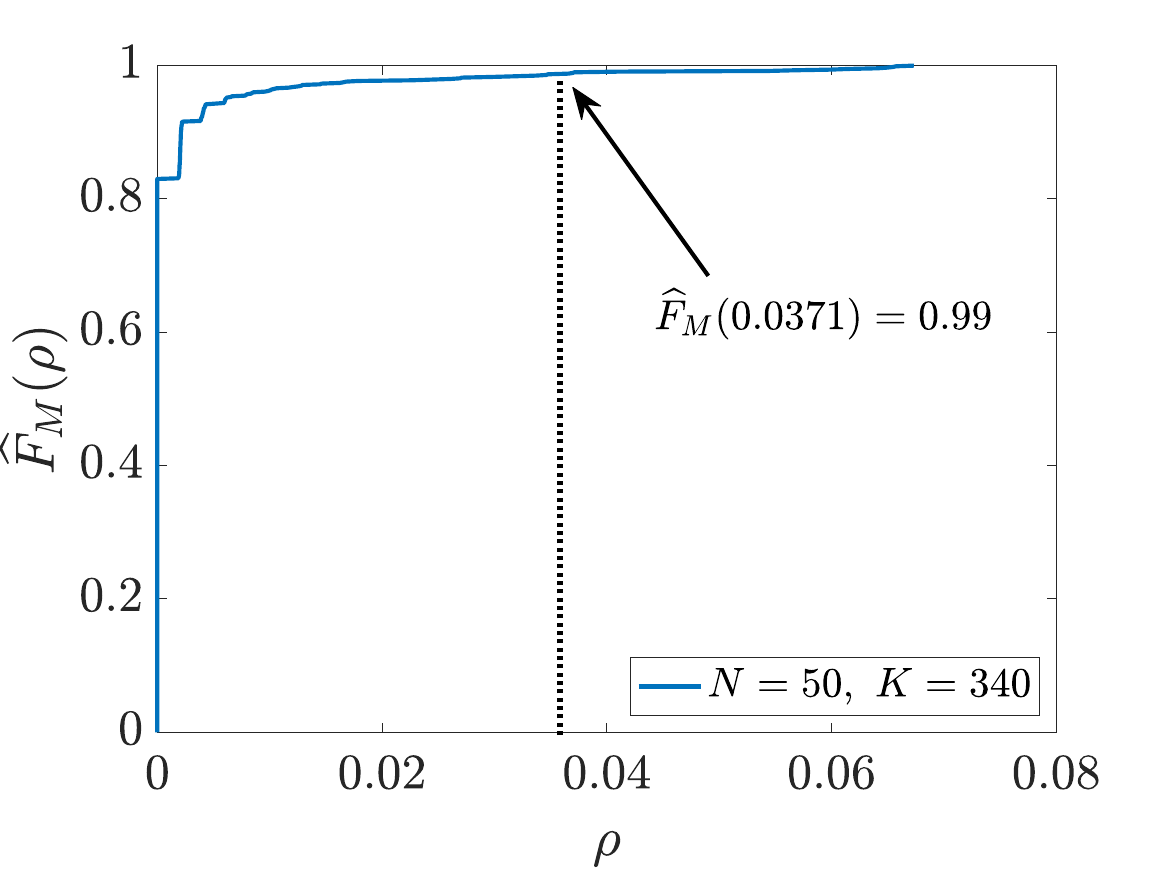}%
\caption{$\widehat{F}_{M}$ for $N=50$ and $\nu = 0.25$; see Fig.~\ref{fig:fHatVsNuN50}b. For small graphs, where the adversarial perturbation removes a significant number of edges, our theory, \eqref{eq_Ki} prescribes $K = 340$ for $\epsilon_{\rm tol} = 0.04$, which is sufficient for this case.}
\label{fig:FhatN50_025}
\end{figure}

For $N = 50$, we observe a second error mode in the empirical PDF in Fig.~\ref{fig:fHatVsNuN50}a when $\nu = 0.1$. The error mode persists across all values of $K$. This error mode corresponds to edges associated with the vertex with the highest eigenvector centrality, which does not sufficiently change across the $K$ repetitions. Setting $\nu = 0.25$, we see in Fig.~\ref{fig:fHatVsNuN50}b that this second error mode vanishes. See Fig.~\ref{fig:FhatN50_025} for the ECDF. From Fig.~\ref{fig:FhatN50_025}, our theory's conservative estimate of $K$ aligns with the required repetitions needed to achieve a robustness close to (but smaller than) $\rho = 0.05$. Thus, the theory produces estimates that are geared towards more aggressive attacks.

\begin{figure}[!]
	\centering
	\subfloat[Erd\H{o}s-R\'{e}nyi graph]{\includegraphics[width=0.4\textwidth,trim={0 0cm 0 0.5cm},clip]{fHatN1000}%
	}
	
	\subfloat[Barab\'{a}si-Albert graph]{\includegraphics[width=0.4\textwidth,trim={0 0cm 0 0.5cm},clip]{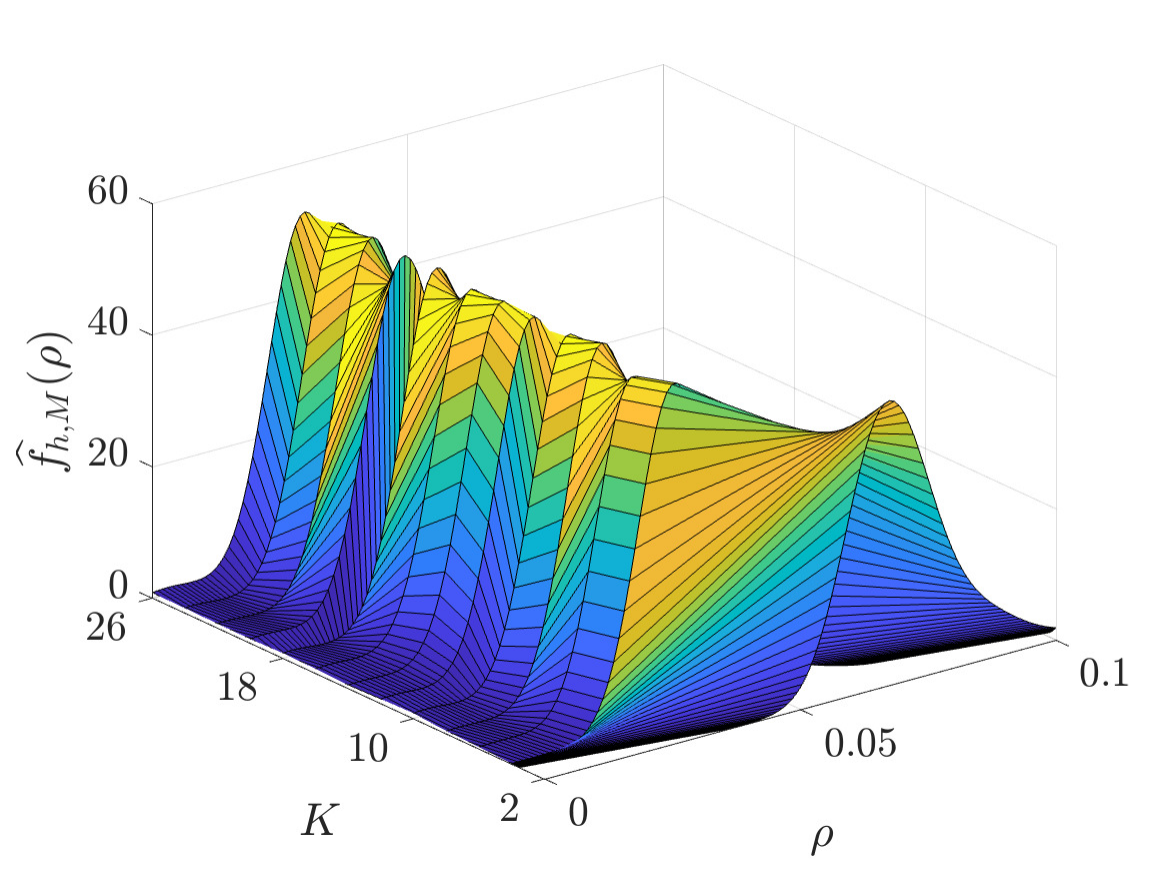}%
	}
	\caption{$\widehat{f}_{h,M}$ for (a) Erd\H{o}s-R\'{e}nyi (ER) versus (b) Barab\'{a}si-Albert (BA) graphs. The settings of the experiment are the same as those used in Fig.~\ref{fig:FHat}. The PDFs associated with the BA graph are flatter and shifted to the right illustrating that BA graphs are more susceptible to attacks that target the network topology like the one considered here.}
	\label{fig:ERvsBA}
\end{figure}

To test the impact of the non-random attack, we'll next consider graphs with preferential attachment, generated using the Barab\'{a}si-Albert (BA) model.  Unlike the case of Erd\H{o}s-R\'{e}nyi (ER) graphs, BA graphs exhibit power law behavior in the vertex degree distributions making them more vulnerable to the non-random component of the attack (removing edges connected to the vertex with the highest eigenvector centrality). As this power law behavior is more pronounced in larger graphs, we'll consider the case $N = 1000$ and compare the effectiveness of the method with those of the ER model shown in Fig.~\ref{fig:FHat}a. The empirical distribution obtained by simulating graphs randomly using the BA model shifts to the right when compared to ER model as shown in Fig.~\ref{fig:ERvsBA}, illustrating the need for a larger $K$ for graphs that have power law degree distributions.

\begin{figure}[!]
	\centering
	\includegraphics[width=0.33\textwidth,trim={0 0cm 0 0.5cm},clip]{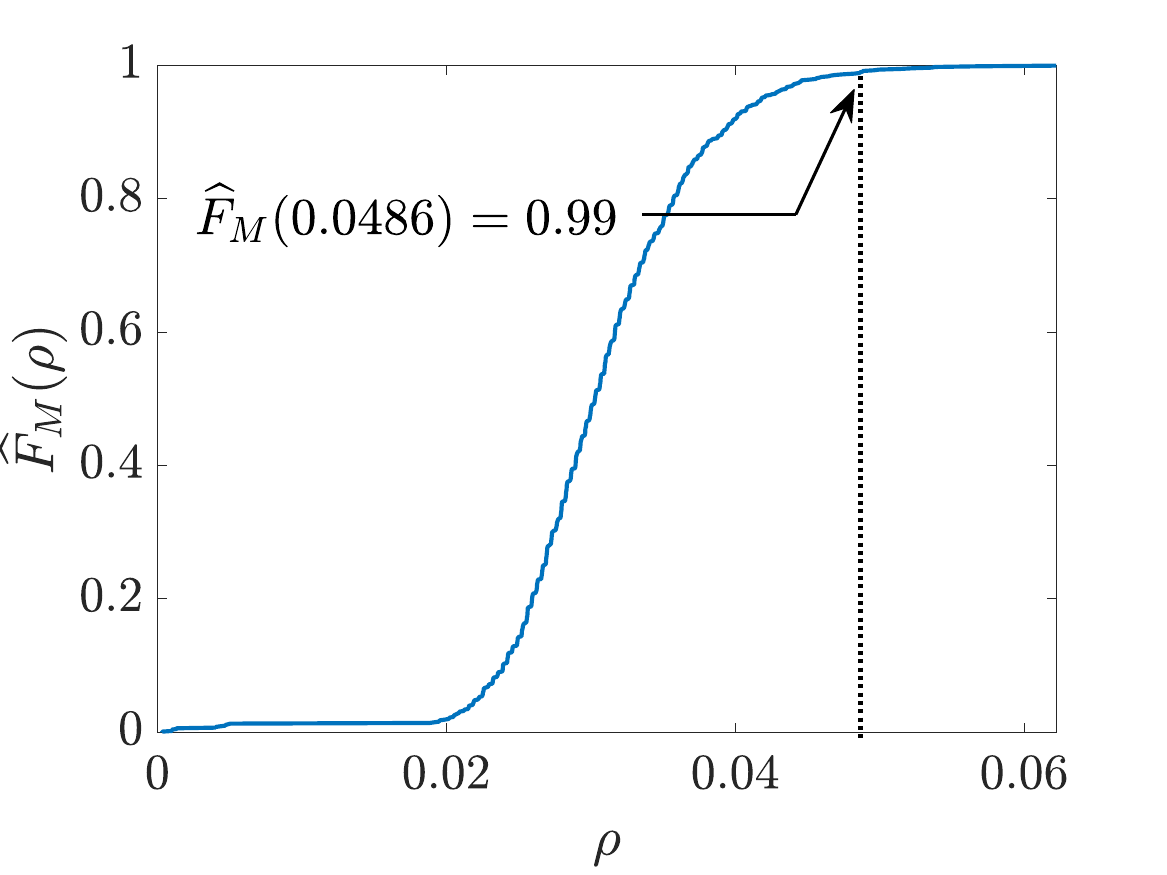}%
	\caption{$\widehat{F}_{M}$ for $N=1000$, $\nu = 0.01$, $K = 10$, and graphs generated with the BA model. }
	\label{fig:BA_FhatN1000_001}
\end{figure}

Finally, Fig.~\ref{fig:BA_FhatN1000_001} shows the ECDF in the BA case for $N=1000$ and $K = 10$. We only needed 10 repetitions for the distribution to converge, well below the theoretical limit given by \eqref{eq_Ki}, and we see that the $\rho = 0.05$ robustness level was achieved with 99\% probability. However, upon comparing this with Fig.~\ref{fig:FHat}a, we see a significant difference between ER graphs and BA graphs. This suggests that a more nuanced approach, which also considers graph topology could be more suitable for scale-free graphs.

\section{Conclusion} \label{sec:conclusion}
For attacks that target graph edges randomly (e.g., \citep{li2020deeprobust}), i.e., do not take the graph topology into account, the defender-assigned perturbation is unnecessary. This results in fewer repetitions needed to correct (potentially) corrupted inputs. In this case, we see that the method also applies to problems where graph data are transmitted across noisy channels, i.e., no attacks. In both of these cases (noise or random attack), one can also transmit the entire signal $\mathbf{t}$ as opposed to the sequence $(\mathbf{t} \oplus \vec{\zeta})|_1,..., (\mathbf{t} \oplus \vec{\zeta})|_K$.  This resembles the case of large graphs investigated in the experiments above, where we found that a small number of repetitions is sufficient for accurate decoding.

For attacks that target specific edges (e.g., \cite{xu2019topology}), the defender-added perturbation $\vec{\zeta}$ becomes necessary. It serves the role of concealing the graph structure from potential attackers. The bound on the number of repetitions derived above tends to be robust to more aggressive attacks. This is understandable as it aims to be robust to attacks of unknown nature. For scale-free graphs, we observe that larger numbers of repetitions are needed to achieve accurate decoding with very high probability (99\%). This suggests that extending the present approach to incorporate elements of network structure could be an effective error correction mechanism for scale-free networks. 

\textit{Acknowledgement}. This work was supported in part by the NYUAD Center for Interacting Urban Networks (CITIES), funded by Tamkeen under the NYUAD Research Institute Award CG001, and in part by the NYUAD Research Center on Stability, Instability, and Turbulence (SITE), funded by Tamkeen under the NYUAD Research Institute Award CG002. The views expressed in this article are those of the authors and do not reflect the opinions of CITIES, SITE, or their funding agencies.

\bibliographystyle{apsrev4-2}
\bibliography{references}
	
\end{document}